\documentclass[a4paper,11pt]{article}
\usepackage{jheppub} 
\usepackage[T1]{fontenc} 
\usepackage{amsfonts}
\usepackage{soul}
\usepackage{amsmath}
\usepackage{amssymb}
\usepackage{amsthm}
\usepackage{array}
\usepackage{axodraw}
\usepackage{booktabs}
\usepackage{color}
\usepackage{comment}
\usepackage{graphicx}
\usepackage{multirow}
\usepackage{slashed}
\usepackage{subfigure}
\usepackage{theorem}
\usepackage{textcomp}

\numberwithin{equation}{section}

\newcommand{\be}{\begin{equation}}
\newcommand{\ee}{\end{equation}}
\newcommand{\beq}{\begin{equation}}
\newcommand{\eeq}{\end{equation}}
\newcommand{\bea}{\begin{eqnarray}}
\newcommand{\eea}{\end{qnarray}}

\def\<{\left\langle}
\def\>{\right\rangle}

\definecolor{brickred}{RGB}{150,25,14}

\def\ptmis{\not\!\!P_T}

\title{
\hfill {\small SHEP-13-05}
\\
What Does the CMS Measurement of W-polarization Tell Us about the Underlying Theory
 of the Coupling of W-Bosons to Matter?
}

\author[a,b]{Alexander Belyaev}
\author[a]{Douglas Ross}
\affiliation[a]{School of Physics \& Astronomy, University of Southampton,
        Highfield, Southampton SO17 1BJ, UK}
\affiliation[b]{Particle Physics Department, Rutherford Appleton Laboratory,
       Chilton, Didcot, Oxon OX11 0QX, UK}
\emailAdd{a.belyaev@soton.ac.uk}
\emailAdd{d.a.ross@soton.ac.uk}

\abstract{
We discuss  results of the CMS collaboration
on the sensitivity of the LHC to $W$ boson polarisation 
in the process  $pp\to W^\pm + jet \to e^\pm jet+\ptmis$
using the $L_P$ variable directly connected to $\theta^*$ 
angle of the outgoing lepton in the rest frame of the decaying $W$.
We have shown that
for a given  $L_P$, interference between different polarizations of the  $W$-boson  is  not negligible, and
needs to be taken into account when considering the 
differential cross-section
with respect to $L_P$.
The  $L_P$ variable suggested by CMS collaboration
is highly suitable  variable
to study LHC sensitivity to $g_V,g_A$ couplings of $W$-boson to fermions.
We note that   the  experimental  sensitivity
to W-boson polarization which is  much higher than that
to  ($g_V,g_A$) parameter space can be turned around and used
to identify deviations from the Standard Model as a signal for  new physics at the LHC.
}

\begin{document}
\maketitle
\flushbottom
\tableofcontents
\section{Introduction}

The CMS collaboration \cite{Chatrchyan:2011ig} has performed a measurement of the
distribution of positive, $(f_+)$, negative, $(f_-)$, and longitudinal, $(f_0)$,
polarisations of the $W$-boson in the process
\beq p\,  p \ \to \ \mbox{jet } \, +
\, W^{\pm} \ \to \ \mbox{jet } \, \ell^{\pm} \, \nu  \label{proc1}
\eeq
for $\ell^\pm=e^\pm \mbox{ and } \mu^\pm$.
Such an
analysis is possible if one can measure the polar, $\theta^*$ and azimuthal,
$\phi^*$ angles of the outgoing lepton in the rest frame of the decaying $W$
(relative to an axis defined by  the direction of the $W$ in the incoming CM
frame).

Since the experiment only observes the decay products of the $W$ in the laboratory
 frame (as opposed to the rest-frame of the $W$), one cannot, in general, neglect the
effect of interference between the production/decay amplitudes for a $W$ of different polarization.
In other  words, the polarization of the $W$-boson is {\it not} an observable. Nevertheless,
in terms of the angles $\theta^*, \phi^*$  defined above the differential cross section (in the case
of $W^-$) may be written  as
\begin{eqnarray} \sigma \frac{d^2\sigma}{d\cos\theta^* \, d\phi^*} & = &
 \frac{1}{4\pi} \Bigg\{
f_+ \frac{\left(1-\cos\theta^*\right)^2}{2}
+f_- \frac{\left(1+\cos\theta^*\right)^2}{2}
+f_0 \left(1-\cos\theta^*\right)^2    \nonumber \\  & & 
+\sqrt{2} g_{+0} \sin\theta^* (1-\cos\theta^*) \cos\phi^*
-\sqrt{2} g_{-0} \sin\theta^* (1+\cos\theta^*) \cos\phi^*
   \nonumber \\  & &
 - g_{+-}\sin^2\theta^*\cos(2\phi^*)
\Bigg\} \label{diffcross} \end{eqnarray}
where $f_i$ are the probabilities for the production and decay of a $W^-$ with
polarization $i$ whereas $g_{ij}$ are the interferences of the amplitudes for production
and decay between $W^-$ with polarizations $i$ and $j$.  We see from eq.(\ref{diffcross})
that after integration over the azimuthal angle, the interference terms vanish and one can
indeed extract the probabilities for the production of the three possible polarizations
from the differential cross-section w.r.t. $\cos\theta^*$.
Here we neglect ``T-odd" contributions at  higher order QCD (see e.g. \cite{Mirkes:1992hu,Bern:2011ie} and references therein).

Unfortunately $\cos\theta^*$ distribution cannot be measured directly.
One needs to find the longitudinal component of the neutrino from $W$-boson decay first
which is not measured but can be deduced form missing transverse momentum, $\ptmis$, the
electron momentum, $p_e$, together with the assumption that the process goes through the
production of an on-shell $W-$boson.
However,  this deduction leads to a quadratic equation with two solutions for
the longitudinal neutrino momentum, $p^\nu_Z$. Therefore   $\cos\theta^*$ cannot in general be measured unambiguously.
 However, the variable  $L_p$ (discussed below),
 proposed in \cite{Chatrchyan:2011ig} which  closely matches  $\cos\theta^*$ for  $W$-bosons
 with the large transverse momentum. \footnote{ The ATLAS collaboration \cite{ATLAS:2012au} has performed a similar analysis using
a different directly measureable variable, $\theta_{2D}$, which matches $\theta^*$  when the transverse momentum of the $W$-boson
dominates its longitudinal momentum.}

Using this variable the analysis of $W-$polarizations has been performed
and the sensitivity of the LHC to $W$-polarization was derived.

We make the point here that since the variable, $L_p$ does not exactly match $\cos\theta^*$, even at very high $p_T(W)$, 
 one should 
 describe the
 differential cross-section $d\sigma/dL_P$ in terms of the sum
 of contributions from the production of $W$-bosons with given helicity
 taking into account  corrections arising from the interference between
 intermediate $W$-bosons of different helicity.

In ref.\cite{Chatrchyan:2011ig}, 
templates were constructed from re-weightings necessary to
produce dependencies on $\cos\theta^*$ from purely
 left-, right-, or longitudinally polarized $W$-bosons, and translating this
 into differential cross-sections in terms of the variable, $L_P$. The 
interference coefficients, $g_{ij}$ were taken to be those for the SM and 
the effect of varying the magnitude of these by $\pm 10 \%$ was investigated and included in the systematic error quoted on the measured values of $f_i$
 \cite{Marrouche:2010fpa}.

In this paper, we take an orthogonal, but complementary approach - namely
we recalculate the differential cross-section with respect to $L_P$ and examine
quantitatively the contribution from the interference terms. We then look at the
effect of altering the vector, $g_V$, and axial-vector, $g_A$, couplings
of the $W$-bosons to quarks, in order to investigate the extent to which
the results quoted in ref.\cite{Chatrchyan:2011ig} can be used to identify deviations from the Standard Model as a signal for new physics in the structure of weak interactions. 
In our analysis, we do not
 extract values for the $f_i$ as we cannot disentangle the interference
contributions.
 We find that for any
of the partonic sub-processes contributing to the process (\ref{proc1}), there
is indeed a sizeable interference effect, but for reasons that must be
coincidental (the three partonic sub-processes are folded with independent
parton-distribution functions), the overall effect is suppressed owing to the fact that
the dominant partonic sub-process is the one for which the interference is smallest.
Nevertheless, the remnant
interference effect is of the order of the  {statistical} experimental errors quoted in
ref.\cite{Chatrchyan:2011ig} and presumably significantly larger than the
current error bars that would be extracted from an analysis over the current (and future)
entire integrated luminosity. 
{For future analysis of $W$-boson helicity distributions, with improved 
statistics, it would be advisable to account directly for interference, or
 alternatively to use templates in which the inteference  coefficients,
 $g_{ij}$, are also  parameters to be fitted.}

Since interference cannot be neglected,
{rather than extracting the probabilities, $f_i$, one should simply
 parameterize the weak interactions in terms of the couplings, $g_V$ and $g_A$}
We  show that the differential
cross section $d\sigma/dL_P$ is not very sensitive to changes in $g_V$ and/or $g_A$,
but rather the 68\% and 95\% CL contours engulf a rather large region in the
 $g_V \, - \, g_A$ plane.
Thus, although the results reported in \cite{Chatrchyan:2011ig} 
{lead to a measurement of} the probabilities $f_i$
to a high degree of accuracy, we find that the accuracy to which the vector and axial vector couplings are measured is considerably poorer.
We emphasize that it is the accuracy to which the LHC can determine these couplings,
rather than the distribution of polarizations of the $W$-boson, that provides direct sensitivity
to the underlying theory (i.e. sensitivity to deviations from the Standard Model).

 The values of $g_V$ and $g_A$ have previously been extracted from
weak decays at low energies and shown to be consistent with the Standard Model
 (SM). {
 If there is new physics beyond the standard model, then the effect of such new physics
 on the effective values of these couplings is likely to be energy dependent, so that it is
informative to compare the values extracted at low-energy weak interactions with those measured
for on-shell W-bosons.}

In this paper we plot the domain
of $g_V$ and $g_A$ couplings extracted from the angular distributions in $W$-boson
 production and decay at the LHC and show that whereas it is compatible with the
SM result, there remains a considerable range for the ratio $g_A/g_V$
 which is still compatible with data.

This paper is organised as follows.
In Section 2 we provide details on the $W$-boson interference
contributing to $d\sigma/dL_P$ and estimate LHC accuracy to measure  the vector and axial vector couplings. In section 3 we draw our conclusions.

\section{LHC sensitivity to the process $pp\to W^\pm + jet \to e^\pm + jet+\ptmis$ }

\subsection{Setup for our analysis}
To explore the LHC sensitivity  for the process  $pp\to W^\pm + jet \to e^\pm + jet+ \ptmis$,
we have evaluated the helicity amplitudes for this process
and have created the respective  Monte-Carlo parton-level
generator linked to PYTHIA~\cite{PYTHIA,Sjostrand:2006za} to simulate
effects beyond the parton-level including as well as
effects related to  detector-energy resolution.
This generator is also able to produce events in the Les Houches Accord format~\cite{Alwall:2006yp}
(LHE) which we have used to link to the PGS~\cite{PGS}  fast detector simulation package.

Our study was done at the leading order
and was concentrated on effects related to W-boson polarisation and LHC sensitivity to W-boson couplings to matter.

In our calculation we have used:
\\
1. CTEQ6L PDF
evaluated at the scale $Q^2\, = \, E_T^2(W) \, \equiv \, M_W^2+(p_T(W))^2$,
choosing the renormalisation and factorisation scale equal to each other;
\\
2. $\alpha_s(M_Z)=0.1184$, $\alpha_{em}(M_Z)=0.007818$.
To be specific, in our analysis we have concentrated on the $e^-$ channel
and have used kinematic cuts specific for the electron signature.
One should note that an
analysis  for $e^+$ channel as well as for the muon channel
are qualitatively similar and lead to the same qualitative conclusions.

We have taken into account contribution from $q=u,d,s,c$ flavours into
the process $pp\to W^\pm + jet \to e^- jet +\ptmis$ 
which   consists of the following 3 (partonic) subprocesses
for $W^- + jet$ production: \\
a) $q_d \bar{q}_u \to g e^- \bar{\nu}_e$ \\
b) $\bar{q}_u g \to \bar{q}_d e^- \bar{\nu}_e$ \\
c) $q_d g \to q_u e^- \bar{\nu}_e$\\
and  3 analogous subprocesses for $W^+ + jet$ production.

For the analysis we have used the same  set of kinematic cuts
which has been used by CMS collaboration in \cite{Chatrchyan:2011ig}
and which we have summarised in Table~\ref{tab:cuts}.
\begin{table}[htb]
\begin{tabular}{|l|l|}
\hline
\hline
Cut \# 	& 	Cut  value\\
\hline
{\it Cut 1}	& $p_T(W)>50$ GeV \\
{\it Cut 2}	& $p_T(e)>25$ GeV, $|\eta(e)|<2.4$ \\
{\it Cut 3}	& Electron isolation within the cone $\Delta R=0.3$ \\
{\it Cut 4}	& Transverse mass of $W$, $M_W^T, > 50$ GeV  \\\hline
\end{tabular}
\caption{The set of kinematic cuts applied for the process $pp\to W^\pm + jet \to e^\pm + jet + \ptmis$
\label{tab:cuts}}
\end{table}

After the {\it Cut 1} the LO cross section for
the process  $pp\to W^- + jet \to e^- jet+\ptmis$ 
($pp\to W^+ + jet \to e^+ jet+\ptmis$) is equal to 127.3 (202.7) pb which has provided about $4.6\times 10^3$ ($7.3\times 10^3$) events
for 36 pb$^{-1}$  integrated luminosity analysed by CMS in \cite{Chatrchyan:2011ig} @ 7 TeV LHC.
For the  total integrated luminosity about 20 fb$^{-1}$ at the latest 
LHC run @ 8 TeV, the number of expected events is almost 2 orders of magnitude larger providing an excellent framework for improvement of the LHC sensitivity
to $W$-boson polarization and couplings to matter.

\subsection{$L_P$ variable}

To be able to measure polarisation of $W$-boson and its couplings to matter
one should be able to access $\cos\theta^*$ value, which, as we have mentioned above
cannot be extracted unambiguously.
Instead,  ref.\cite{Chatrchyan:2011ig}
have proposed the variable $L_P$ defined by
\beq L_p \ = \ \frac{\mathbf{p_T}(e) \cdot {\mathbf{p_T}(W)}}{|\mathbf{p_T}(W)|^2}, \eeq
where $\mathbf{p_T}(e), \, \mathbf{p_T}(W)$ are the transverse momenta of the
outgoing charged lepton and the $W$ respectively. In terms of the polar and azimuthal
angles $\theta^*, \, \phi^*$ the variable $L_P$ is given by
 \footnote{In this paper we restrict our analysis to the case of a single (narrow)
 jet accompanying the $W$-boson,}
\beq  L_P  \ = \ \frac{1}{2} +\frac{1}{2} \frac{(\hat{s}+M_W^2)}{(\hat{s}-M_W^2)} \cos\theta^*
  \pm \frac{ M_W \sqrt{\hat{s}}}{(\hat{s}-M_W^2)}
\frac{|{\mathbf{p_{||}}(W)|}}{|{\mathbf{p_{T}}(W)|}} \sin\theta^*\cos\phi^*
 \label{lp2} \eeq

where $\hat{s}$ is the invariant square-mass of the $W$+jet and
the related quantity $|\mathbf{p_{||}}(W)|$ is the longitudinal component
of the momentum of the $W$. For sufficiently large $|\mathbf{p_{T}}(W)|$
eq.(\ref{lp2}) may be approximated by
\beq L_P \ \approx \ \frac{1}{2} (1+\cos\theta^*),  \label{approx} \eeq
so that if a  sufficiently large  minimum cut is imposed on $|\mathbf{p_{T}}(W)|$,
the (ambiguous) dependence of $L_P$ on the azimuthal angle $\phi^*$ decouples so that the differential
cross-section w.r.t.  $L_P$  can be used to extract the probabilities for the different $W$
polarisations.

\begin{figure}[htb]
\centerline{\includegraphics[width=0.5\textwidth,angle=-90]{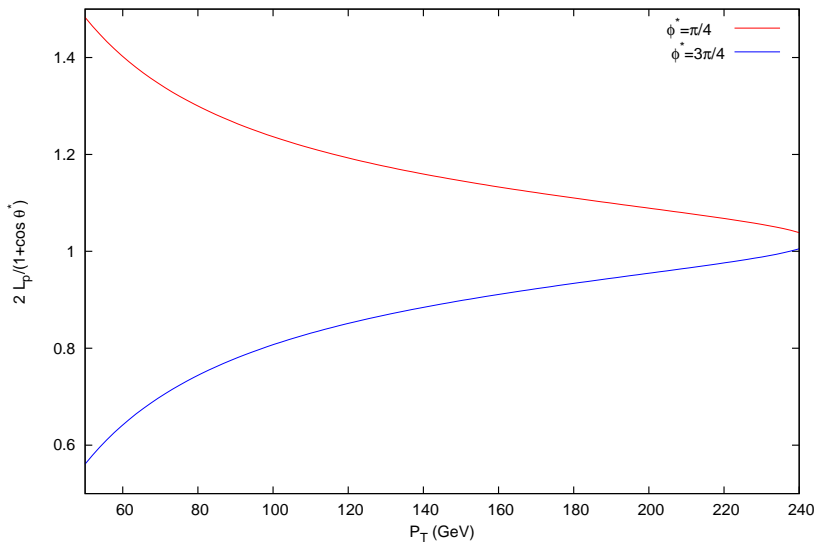}}
\caption{The ratio of $L_P$ to $\frac{1}{2}(1+\cos\theta^*)$ against the
transverse momentum of the $W$ for azimuthal angles $\phi^*=\pi/4$ and
$\phi^*=3\pi/4$. { We have taken $\theta^*=\pi/4$ and
 $\sqrt{\hat{s}} = 0.5$ TeV.} \label{lpvscos}}
\end{figure}

However, it can be seen from Fig.~\ref{lpvscos} that this approximation is only valid
at very high  values of $|\mathbf{p_{T}}(W)|$ and that even for
$|\mathbf{p_{T}}(W)|=240$ GeV
 (close to the kinematic limit for $\sqrt{s}=500$ GeV)
 there is a 5\% difference. In ref.\cite{Chatrchyan:2011ig}, the
cut imposed on $|\mathbf{p_{T}}(W)|$ was 50 GeV, where we can see that
 there are very large corrections to the approximation
 (\ref{approx}). For this figure we have chosen 500 GeV 
 as the partonic subenergy, $\sqrt{\hat{s}}$ since for higher $\sqrt{\hat{s}}$
 the  contribution to the total cross section is below 1\% at 7 TeV LHC.

Consequently, we expect a significant contribution to the cross-section
 $d\sigma/dL_P$ from interference terms - i.e. it is {\it not}
 a good approximation to assume that the dependence on $\phi^*$ factorises in such a way
that the effect of interference on this differential cross-section 
{is small.}

\subsection{The value of the $W$-boson interference for $L_P$ variable}

\begin{figure}[htb]
\includegraphics[width=0.48\textwidth]{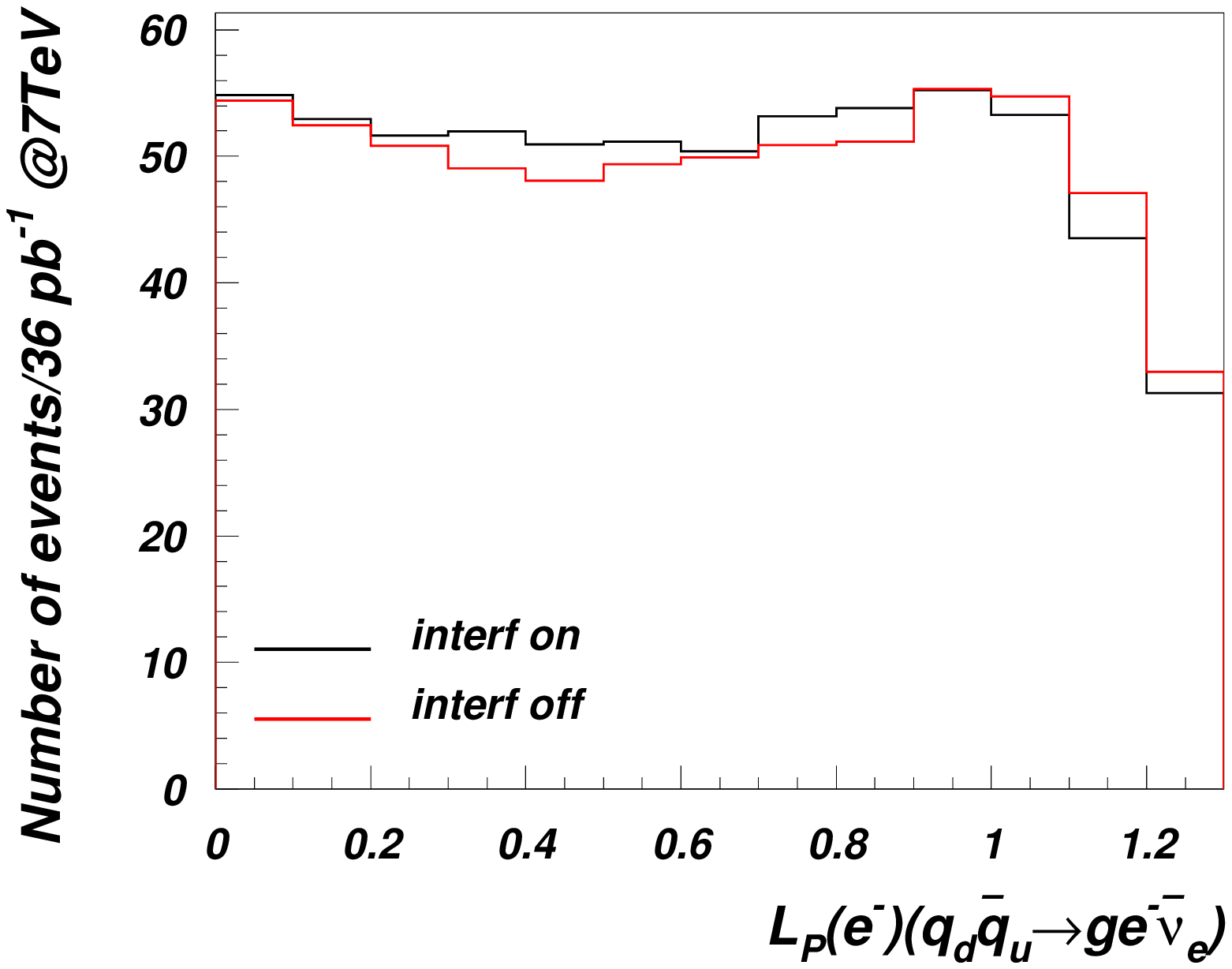}%
\includegraphics[width=0.48\textwidth]{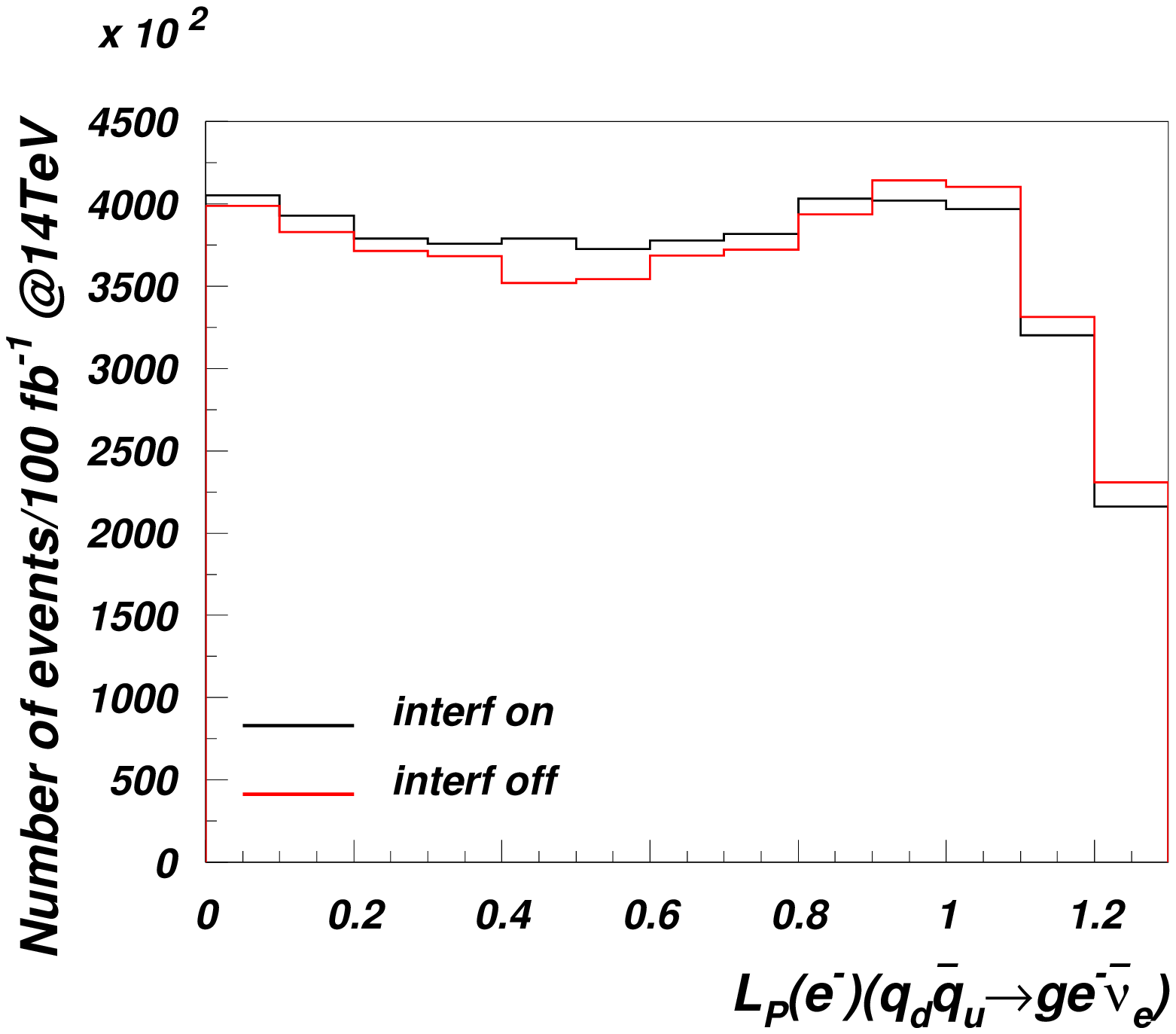}\\
\vskip -1.3cm
\includegraphics[width=0.48\textwidth]{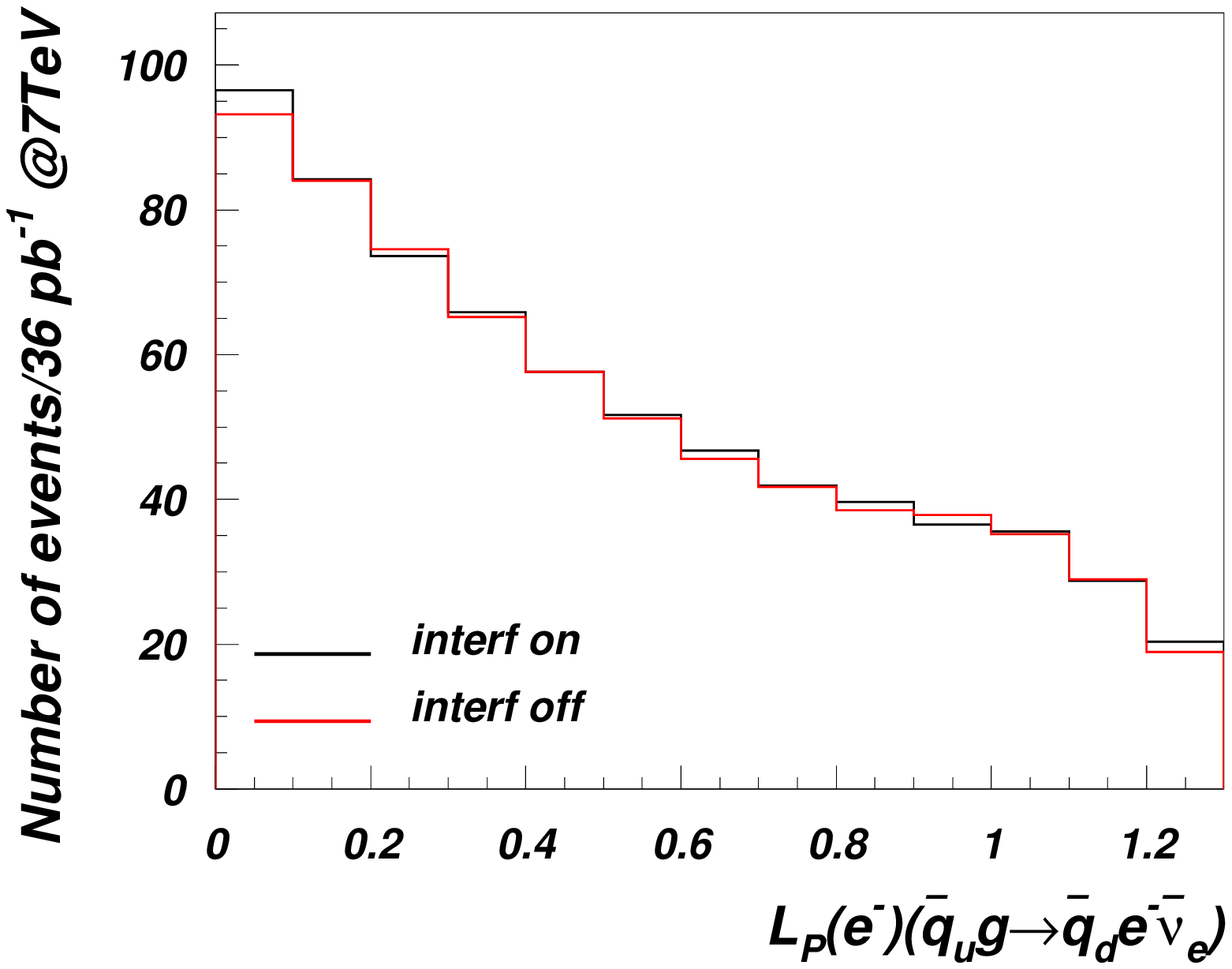}%
\includegraphics[width=0.48\textwidth]{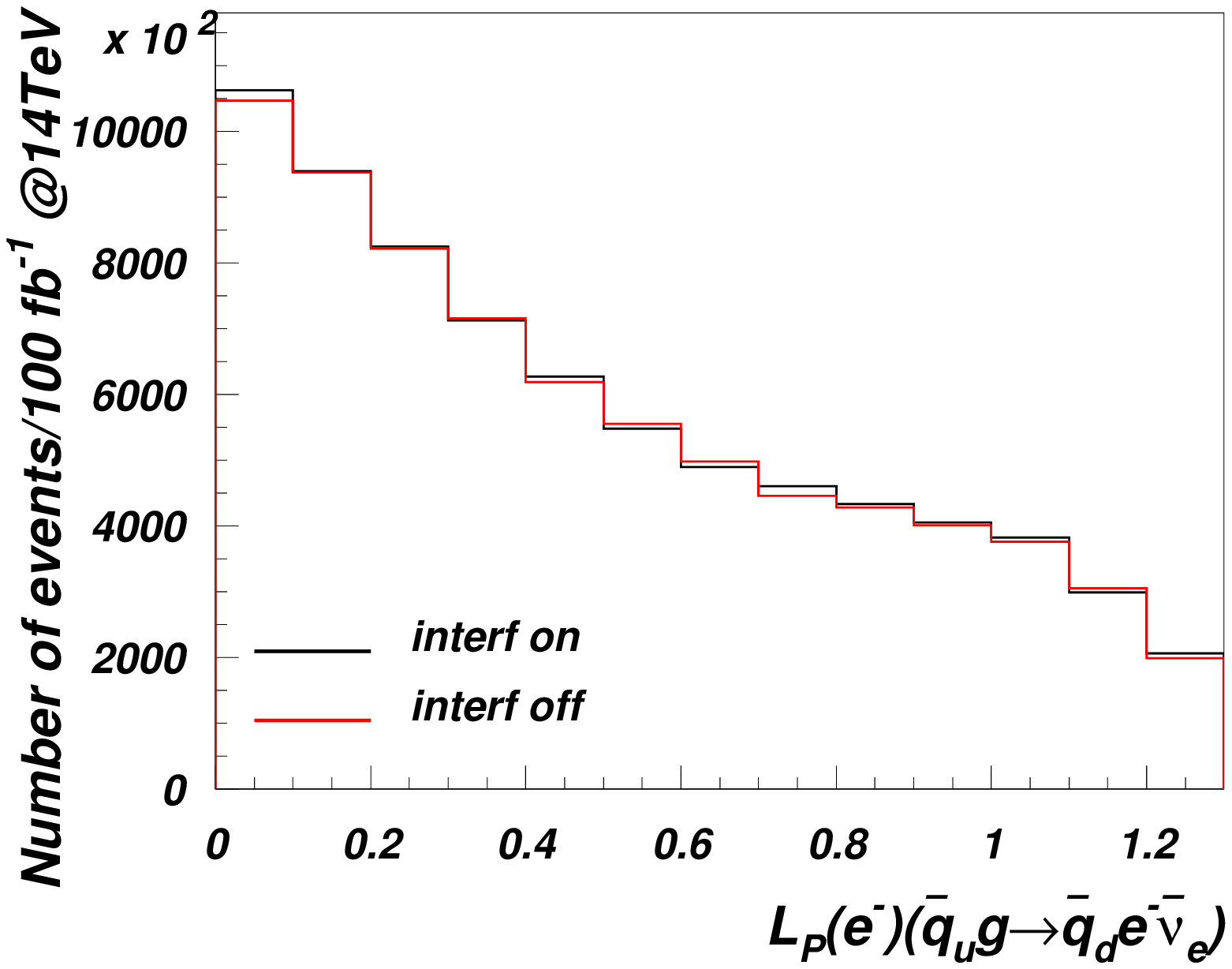}\\
\vskip -1.3cm
\includegraphics[width=0.48\textwidth]{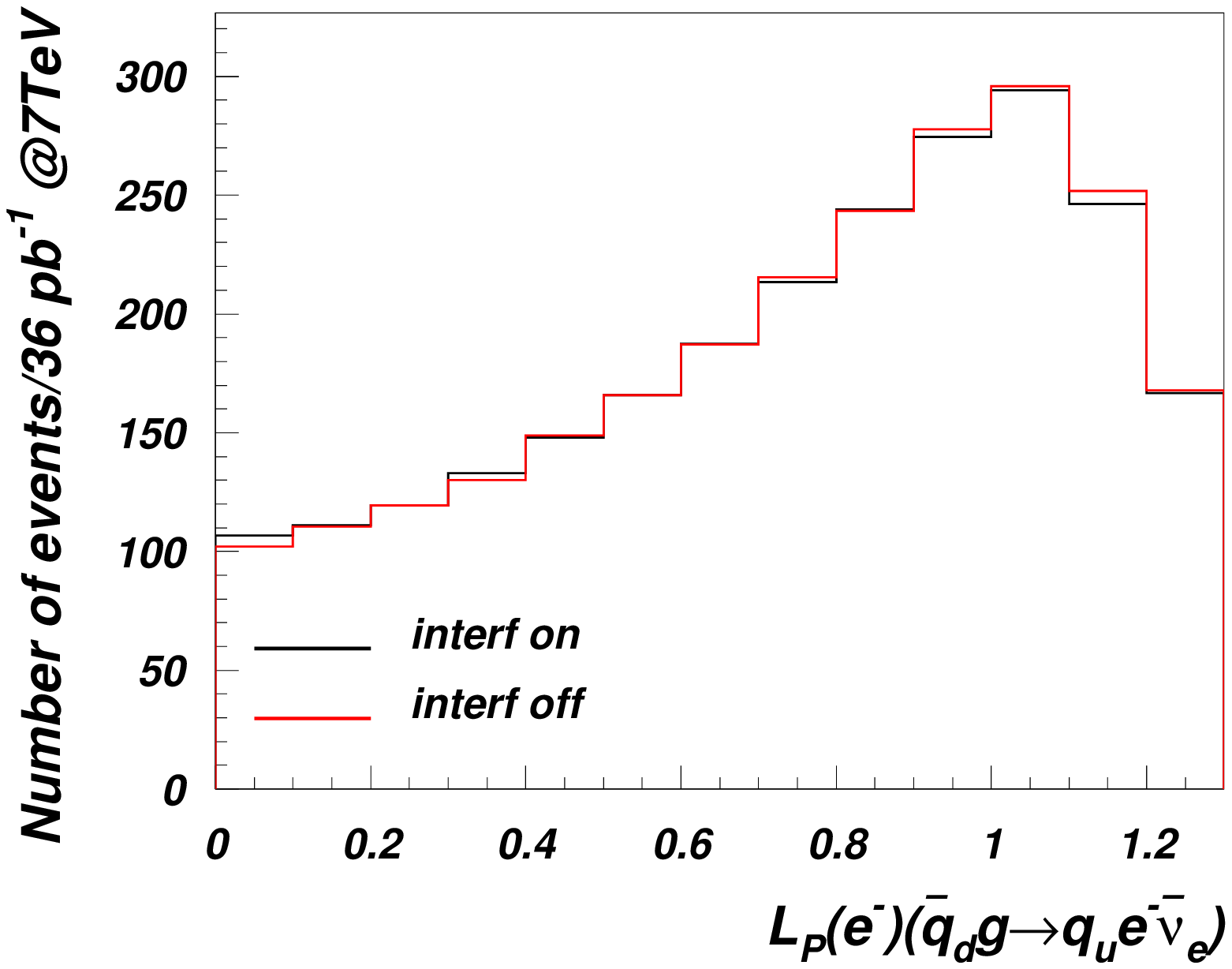}%
\includegraphics[width=0.48\textwidth]{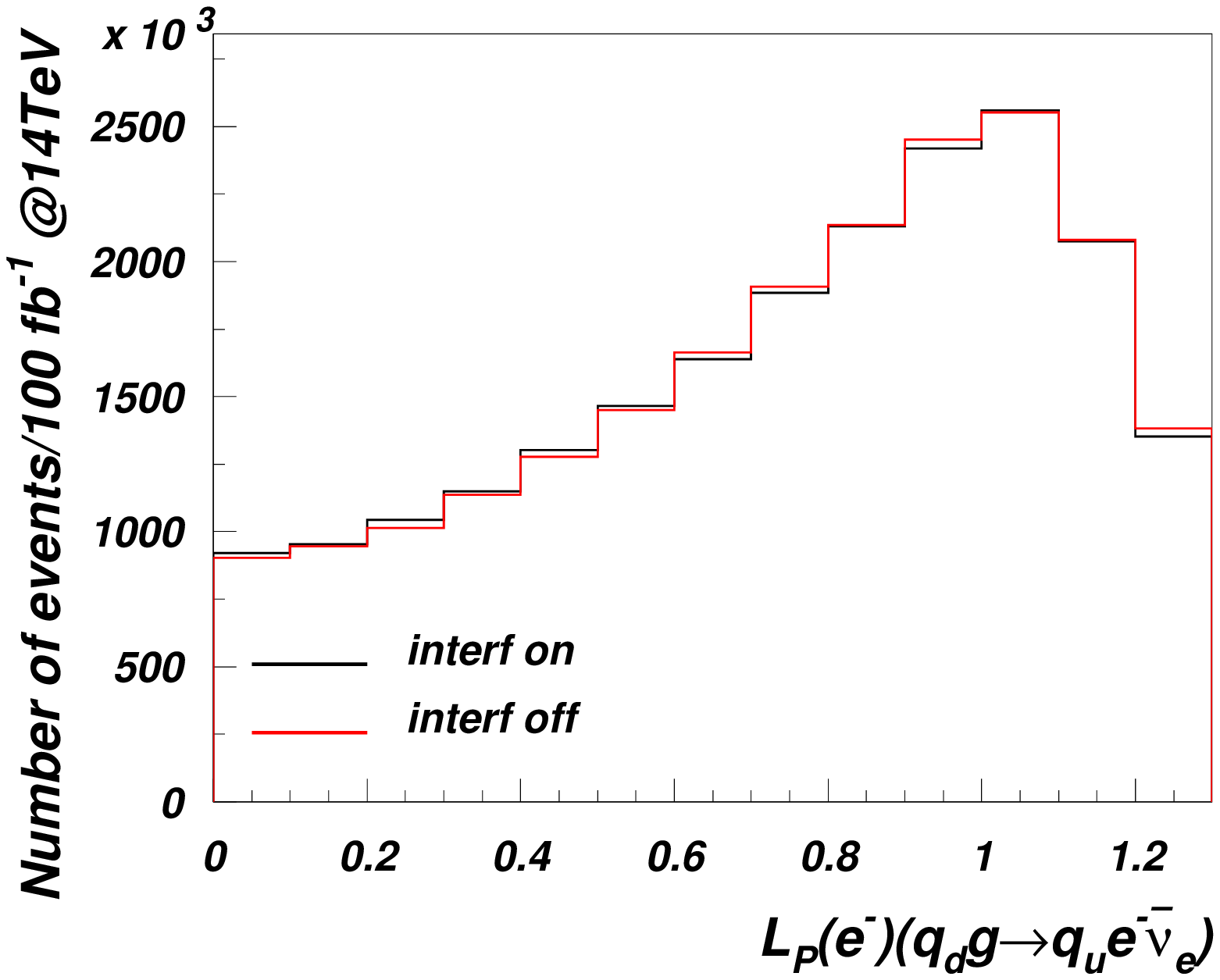}
\vskip -0.5cm
\caption{The effect of the interference of the $W-$ boson polarisations for $L_P$ variable
for individual subprocesses at \@ 7 TeV LHC, 36 pb $^{-1}$ (left column) and
 \@ 14 TeV LHC, 100 fb $^{-1}$ (right column).
Black line -- interference is taken into account, red line -- interference is neglected.
\label{fig:interf-ind}}
\end{figure}

In Fig.\ref{fig:interf-ind} we present $d\sigma/dL_P$ distributions for an individual subprocesses contributing to the process  $pp\to W^- + jet \to e^- jet+\ptmis$  after {\it Cut 1} (see Table~\ref{tab:cuts})
 at \@ 7 TeV LHC, 36 pb $^{-1}$ (left column) and
 \@ 14 TeV LHC, 100 fb $^{-1}$ (right column).
One can see that the relative interference effect
can be above 10\% in the  case of the process
$q_d \bar{q}_u \to g e^- \bar{\nu}_e$ 
and clearly cannot be neglected.

This is the case for both 7 TeV and 14 TeV LHC energies.
At the same time one can see that  the process $q_d \bar{q}_u \to g e^- \bar{\nu}_e$
is not the dominant one; the main contribution to the process $pp\to W^\pm + jet \to e^- jet+\ptmis$ 
actually comes from  the subprocess $q_d g \to q_u e^- \bar{\nu}_e$ .
Therefore, the overall effect of interference in $d\sigma/dL_P$ distribution 
is only at a few percent level as we present in detail below.
So accidentally, the overall interference effect turns out to be quite small, for most of the $L_P$ bins, but nevertheless not negligible.
The situation is qualitatively the same for the process $pp\to W^+ + jet \to e^+ jet+\ptmis$.

\begin{figure}[htb]
\includegraphics[width=0.5\textwidth]{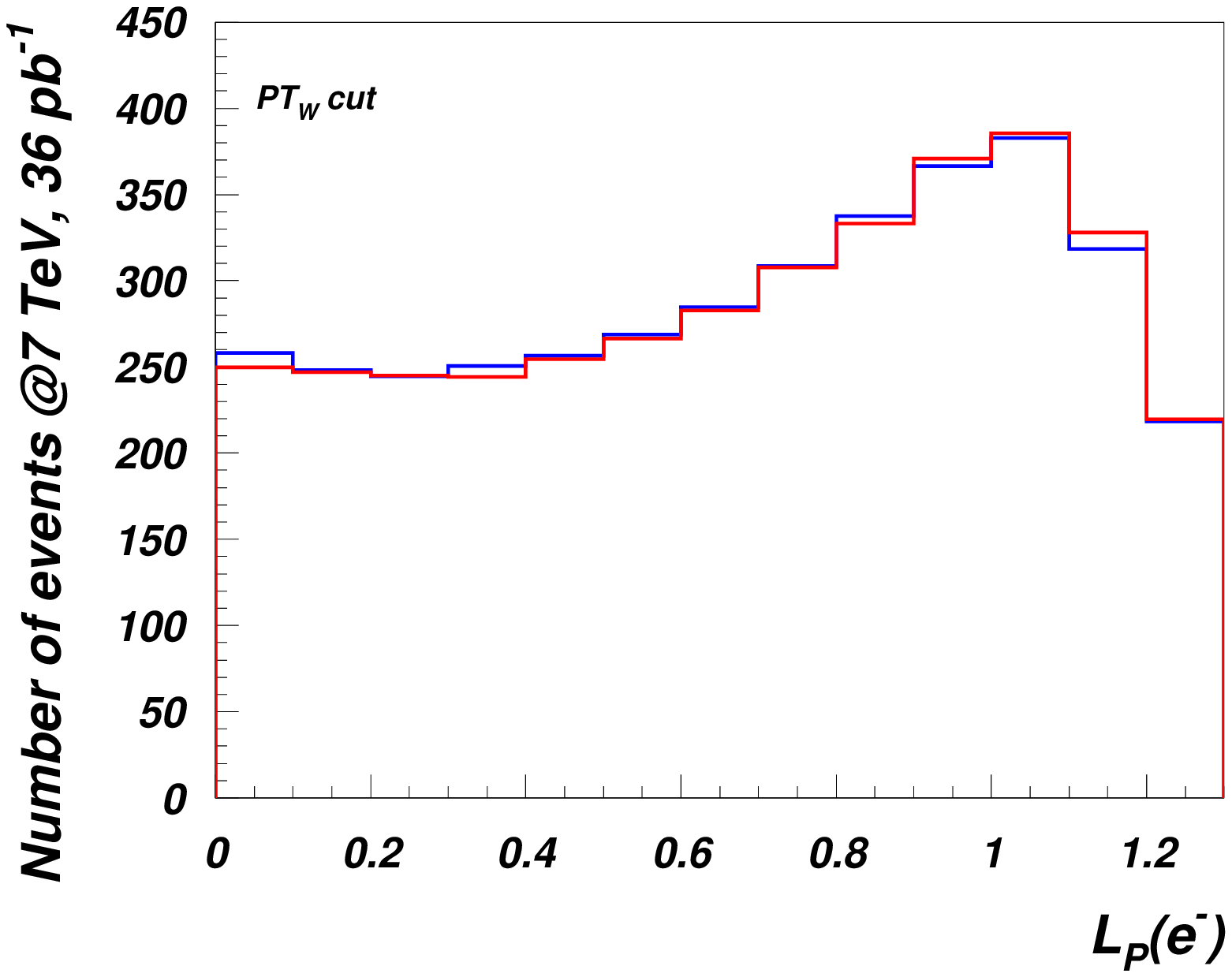}%
\includegraphics[width=0.5\textwidth]{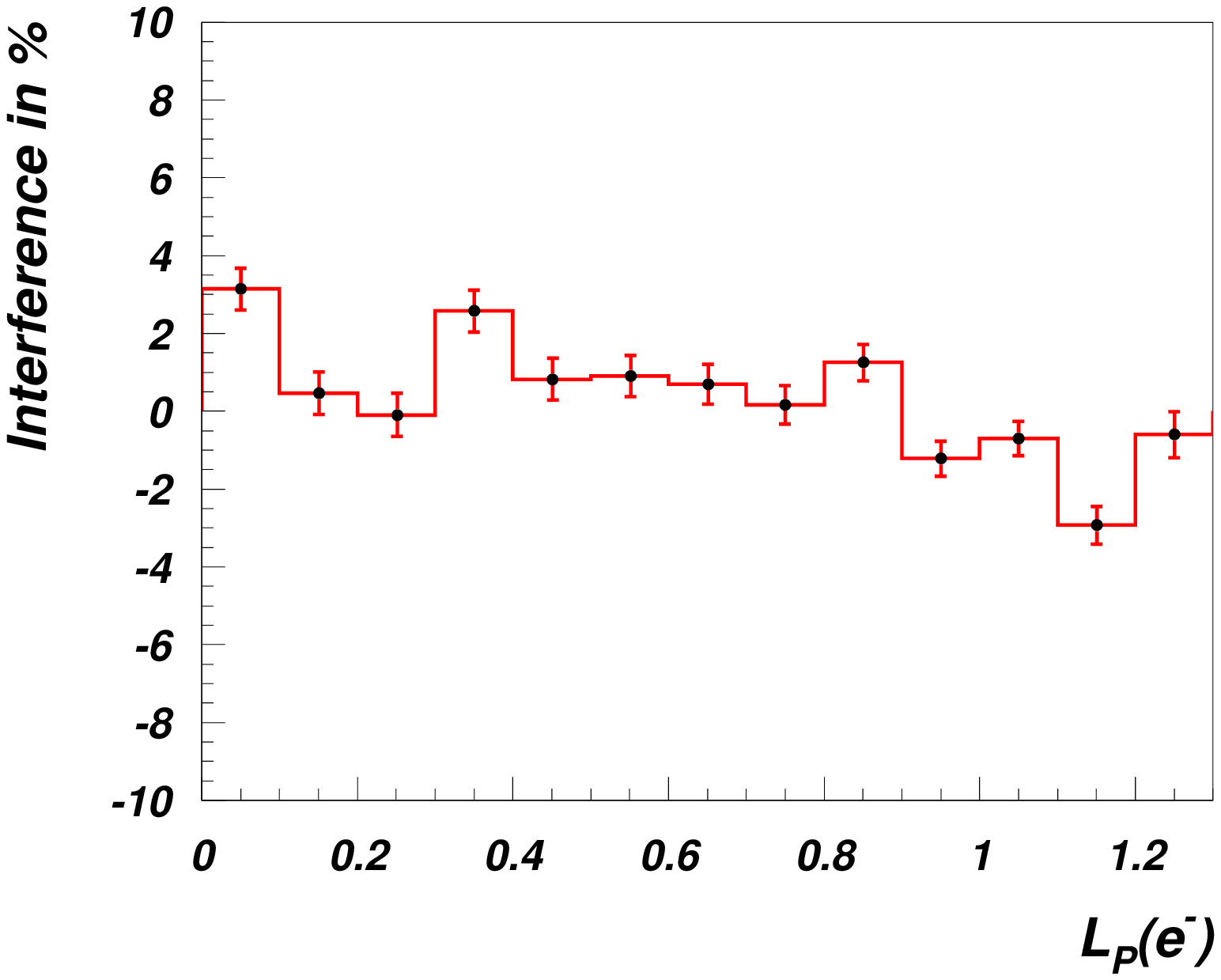}\\
\vskip -1.4cm
\includegraphics[width=0.5\textwidth]{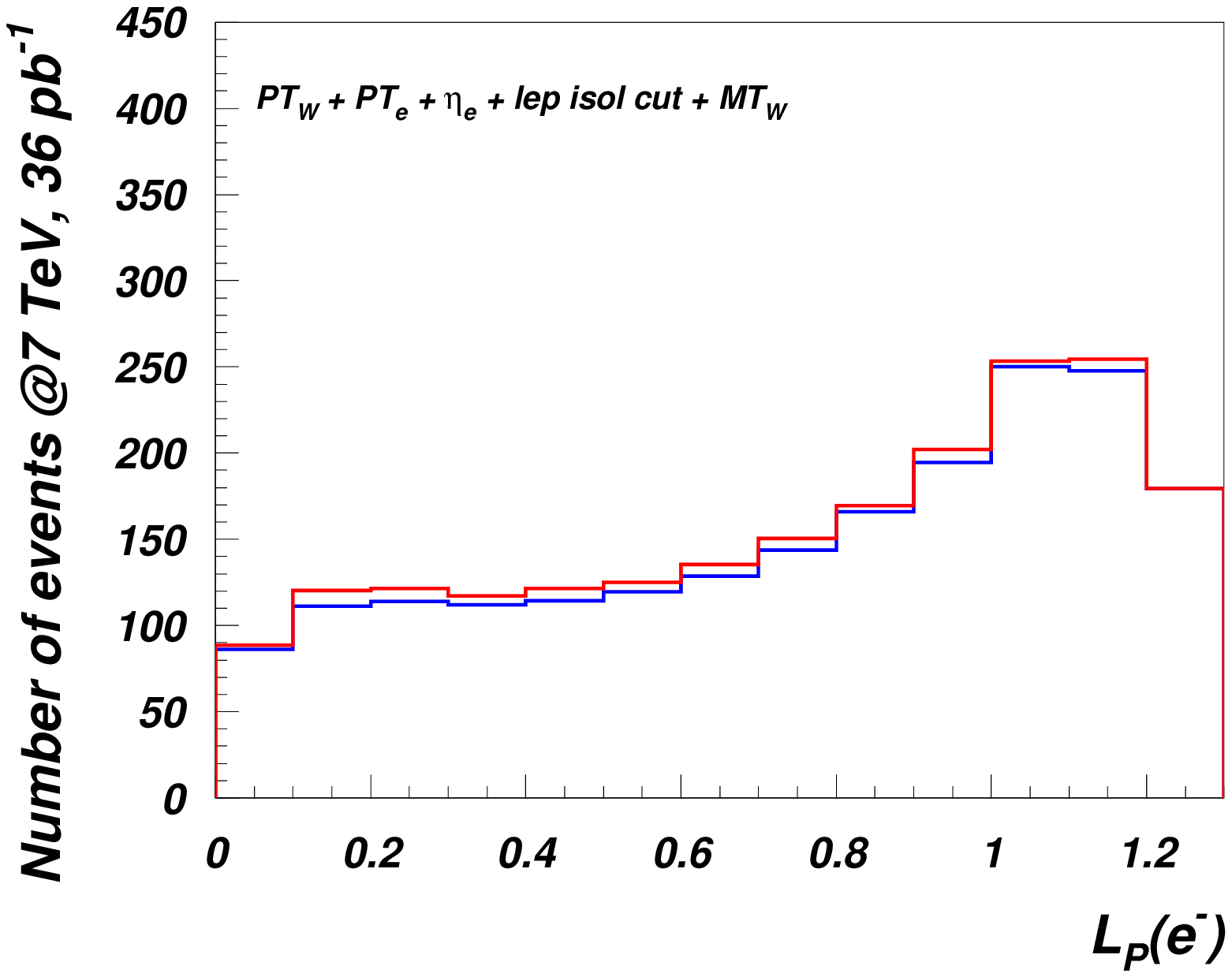}%
\includegraphics[width=0.5\textwidth]{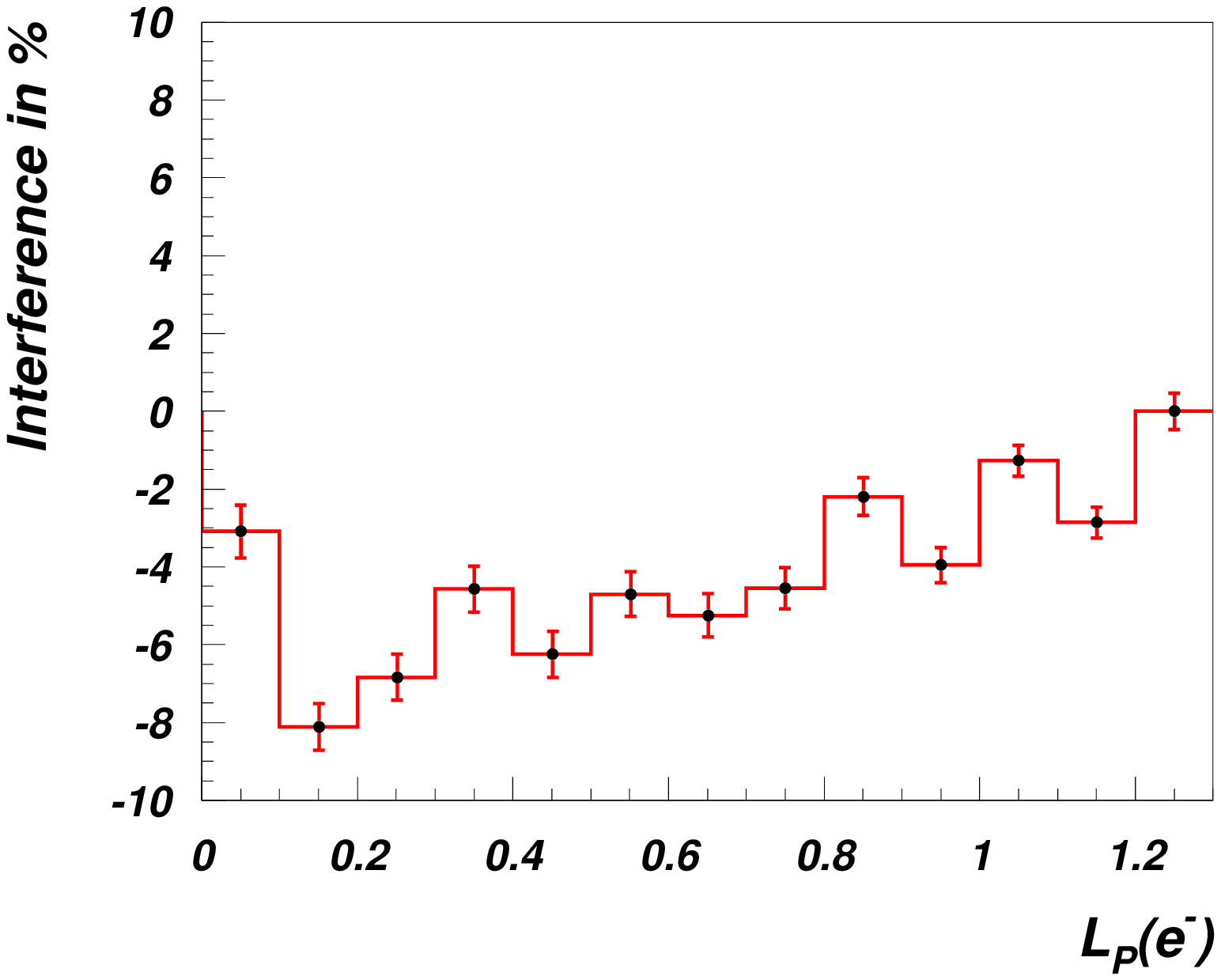}
\vskip -0.5cm
\caption{\label{fig:interf-cuts-}
Overall interference effect of $W$-boson polarisations
in  $d\sigma/dL_P$ \@7 TeV LHC with 36 pb$^{-1}$ after the consequent application of
the kinematic cuts from Table~\ref{tab:cuts} for the process $pp\to W^- + jet \to e^- jet+\ptmis$.
Left: distribution with(blue) and without(red) interference.
Right: the relative value of the interference versus $dL_P$
including Monte-Carlo statistical error.
Upper: distributions after {\it Cut 1},
Lower: distributions after {\it Cuts 1-4}.}
\end{figure}

\begin{figure}[htb]
\includegraphics[width=0.5\textwidth]{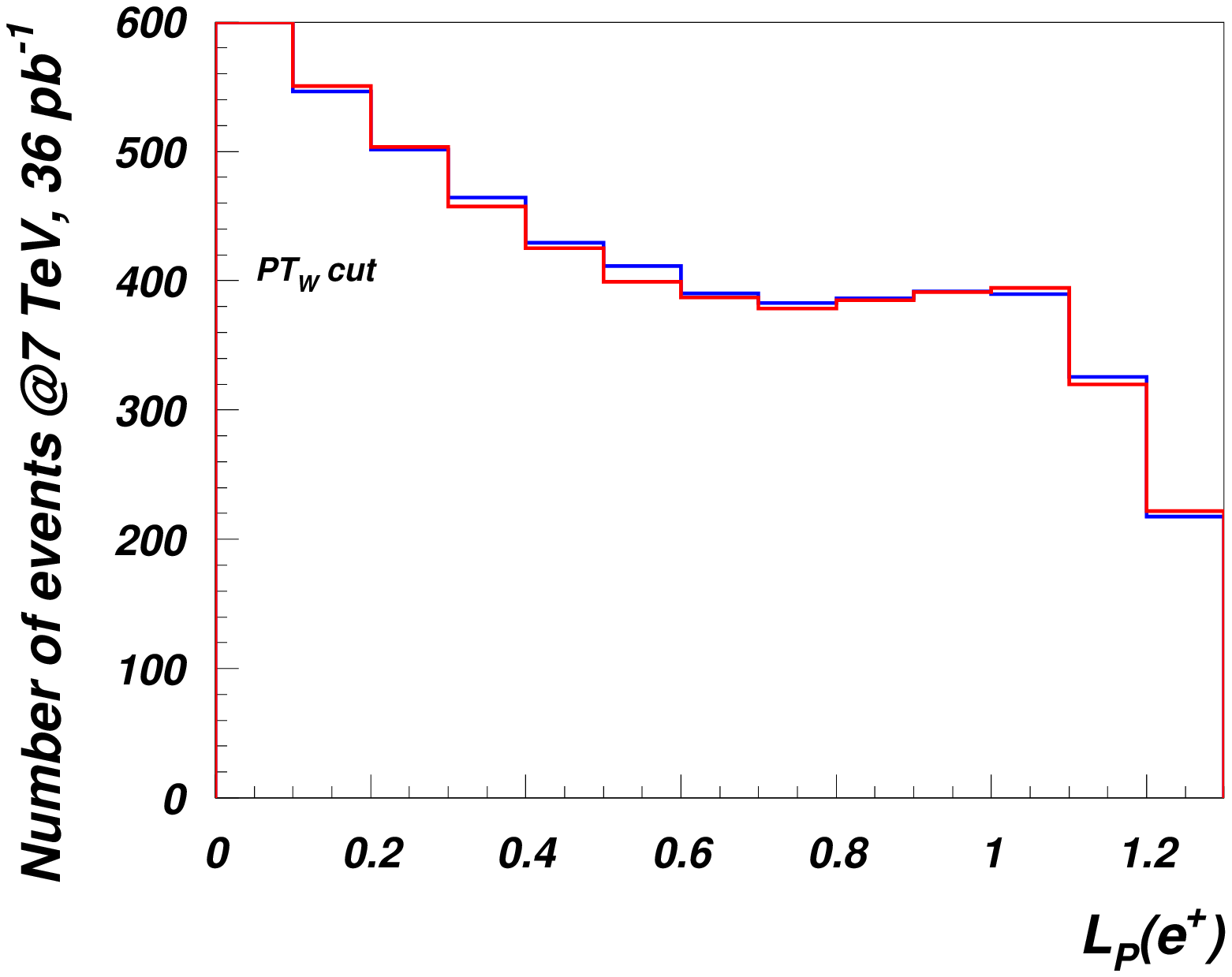}%
\includegraphics[width=0.5\textwidth]{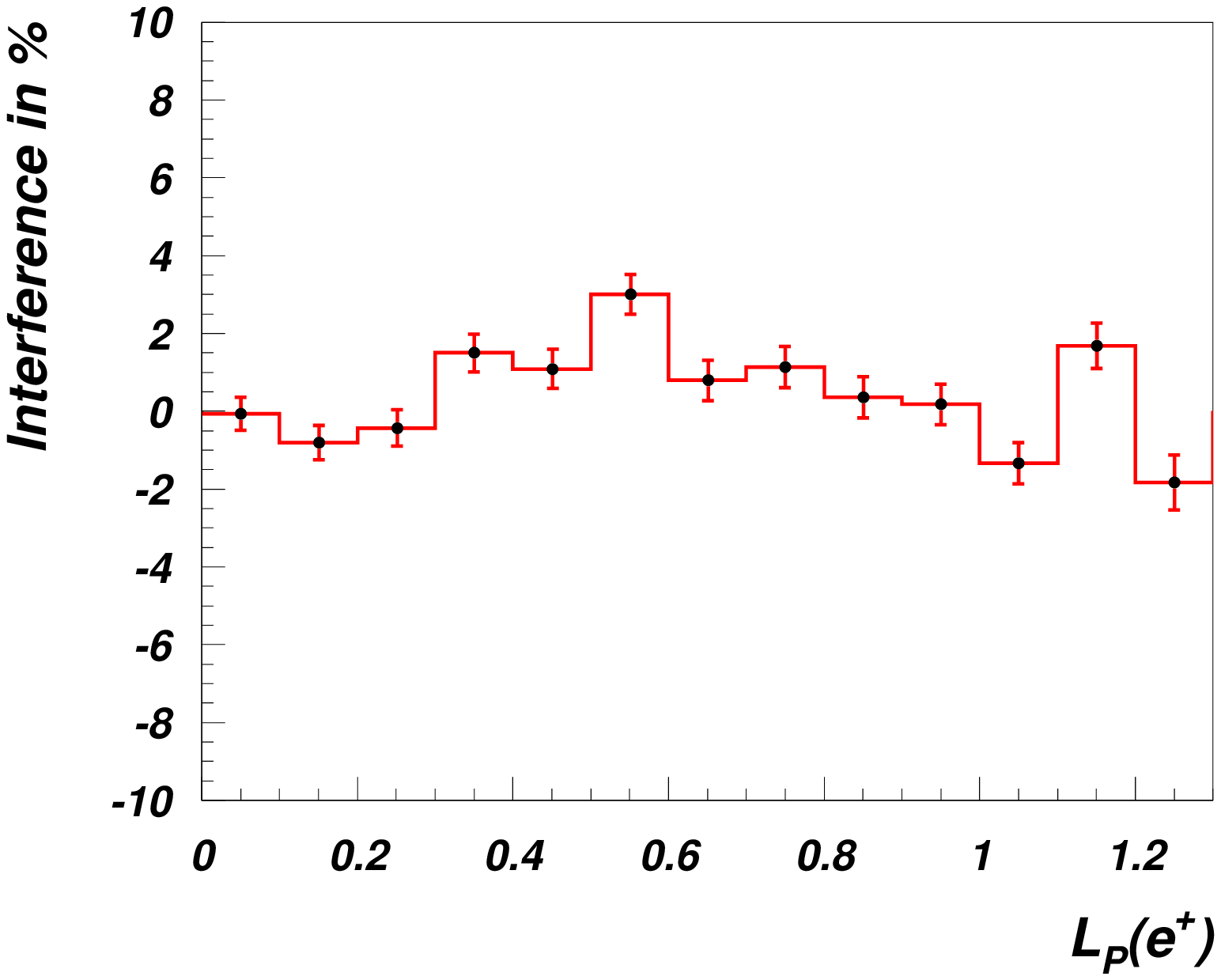}\\
\includegraphics[width=0.5\textwidth]{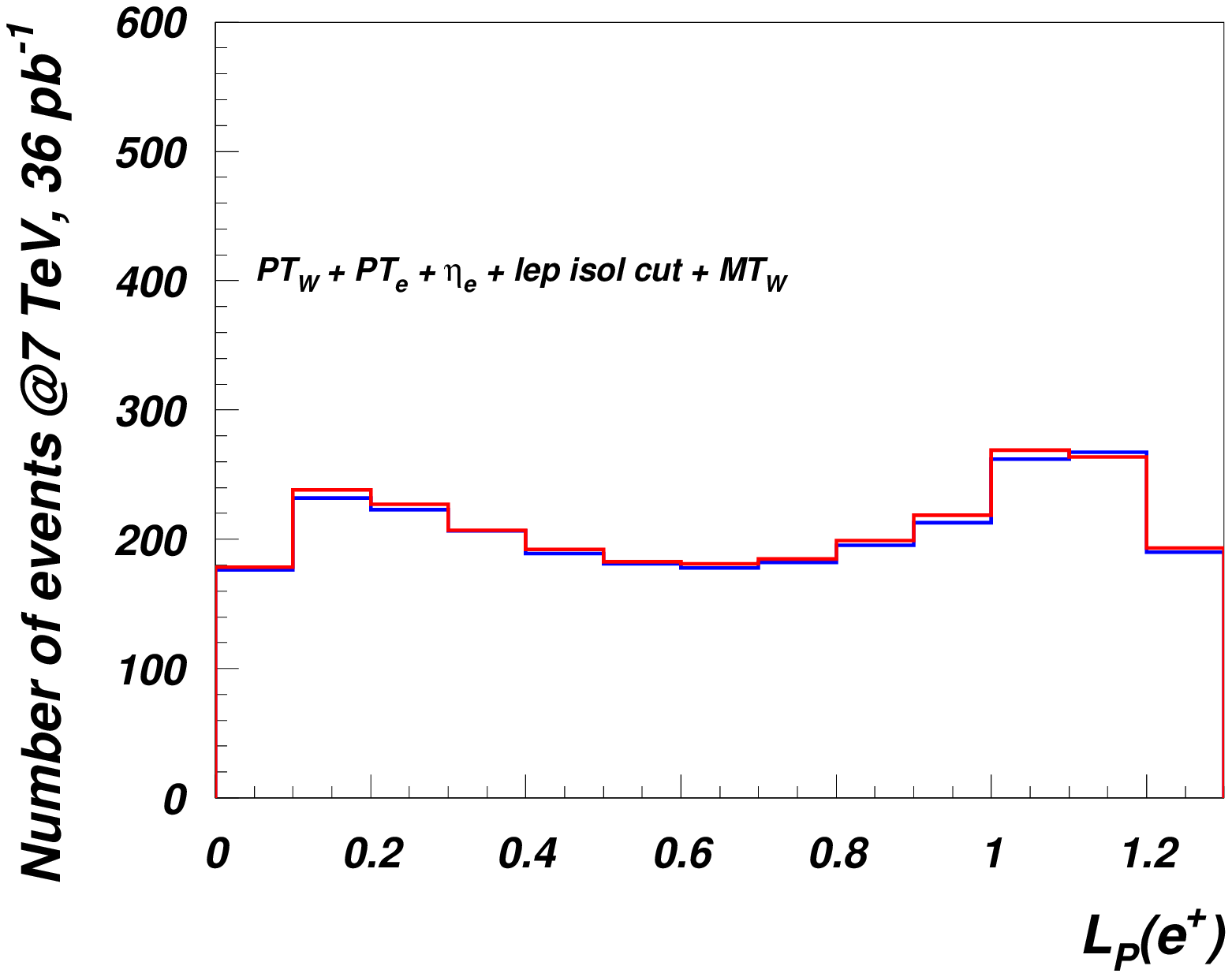}%
\includegraphics[width=0.5\textwidth]{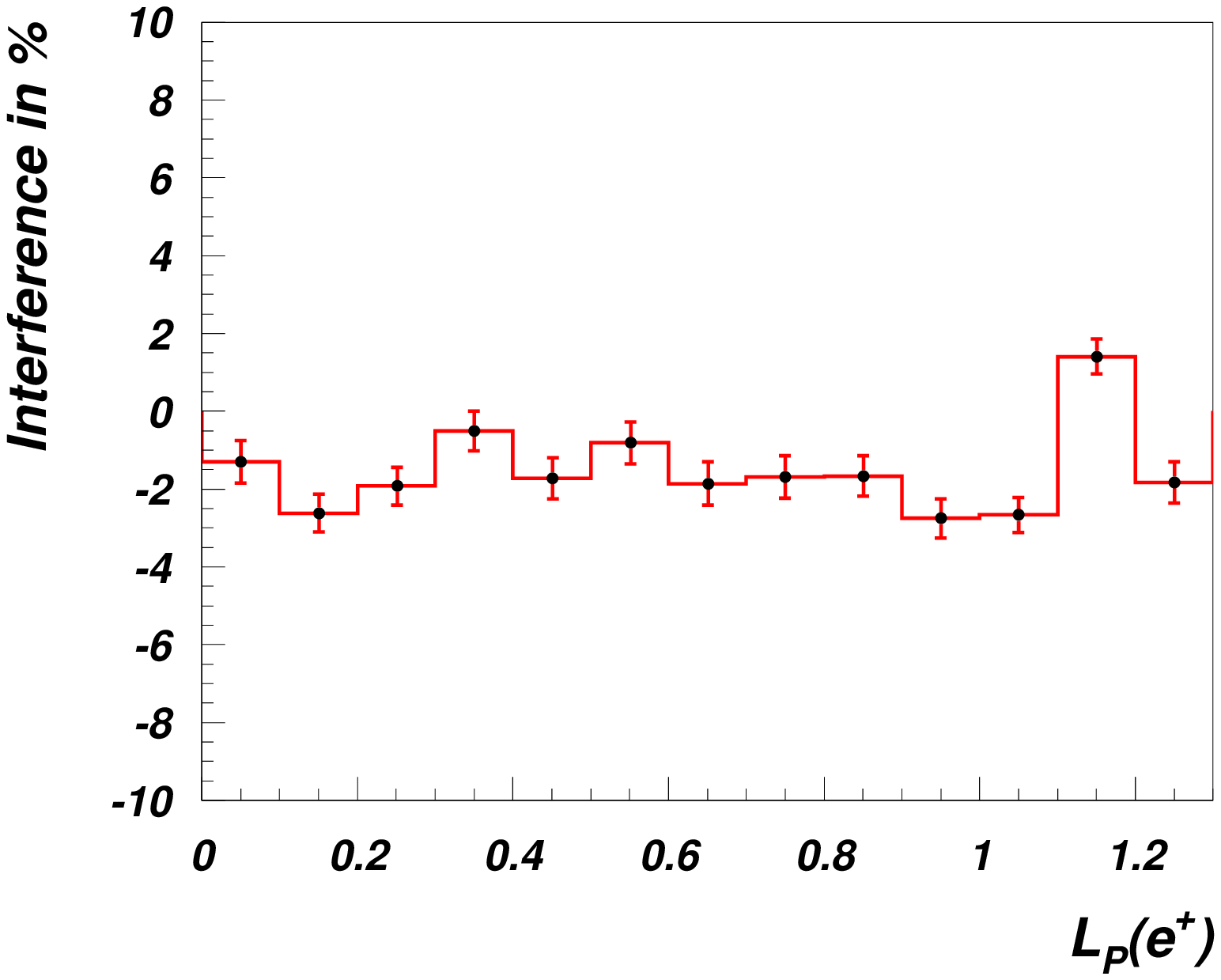}\\
\caption{\label{fig:interf-cuts+}
Overall interference effect of $W$-boson polarisations
in  $d\sigma/dL_P$ \@7 TeV LHC with 36 pb$^{-1}$ after the consequent application of
the kinematic cuts from Table~\ref{tab:cuts} for the process $pp\to W^+ + jet \to e^+ + jets+\ptmis$.
Left: distribution with(blue) and without(red) interference.
Right: the relative value of the interference versus $dL_P$
including Monte-Carlo statistical error.
Upper: distributions after {\it Cut 1},
Lower: distributions after {\it Cuts 1-4}.}
\end{figure}

Let us take a look at the overall interference effect after the consequent application of
the kinematic cuts from Table~\ref{tab:cuts} presented in Fig.~\ref{fig:interf-cuts-}-\ref{fig:interf-cuts-}.
In these figures we present total $d\sigma/dL_P$ distributions at 7 TeV LHC with 36 pb$^{-1}$
for  the processes $pp\to W^- + jet \to e^- jet+\ptmis$ (Fig.\ref{fig:interf-cuts-})
and  $pp\to W^+ + jet \to e^+ jet+\ptmis$ (Fig.\ref{fig:interf-cuts+}) , but
ignoring effects of calorimeter energy smearing, trigger efficiency as well as
effects of jet fragmentation.


One can see here that the  total interference can be as large as about 10\%
in certain bins of $d\sigma/dL_P$ distribution which one can observe from the 
right and middle parts of Fig.~\ref{fig:interf-cuts-} and  Fig.~\ref{fig:interf-cuts+}
respectively.
Another observation is that  { while kinematic cuts suppress overall event rate,
they do not visibly affect the  magnitude of the interference, so that the   relative importance of
such interference  increases as  more and more cuts are applied.}
For low $L_P$ values it reaches -8\% in certain bins for 7 TeV collision.
One should also  mention that for the process $pp\to W^+ + jet \to e^+ jet+\ptmis$   the interference
is smaller than for  $pp\to W^- + jet \to e^- jet+\ptmis$ production.
From this point we will be presenting results only for 
the process
$pp\to W^- + jet \to e^- jet+\ptmis$  recalling that the results for $pp\to W^+ + jet \to e^+ jet+\ptmis$ one are qualitatively the same.

\begin{figure}[htb]
\includegraphics[width=0.5\textwidth]{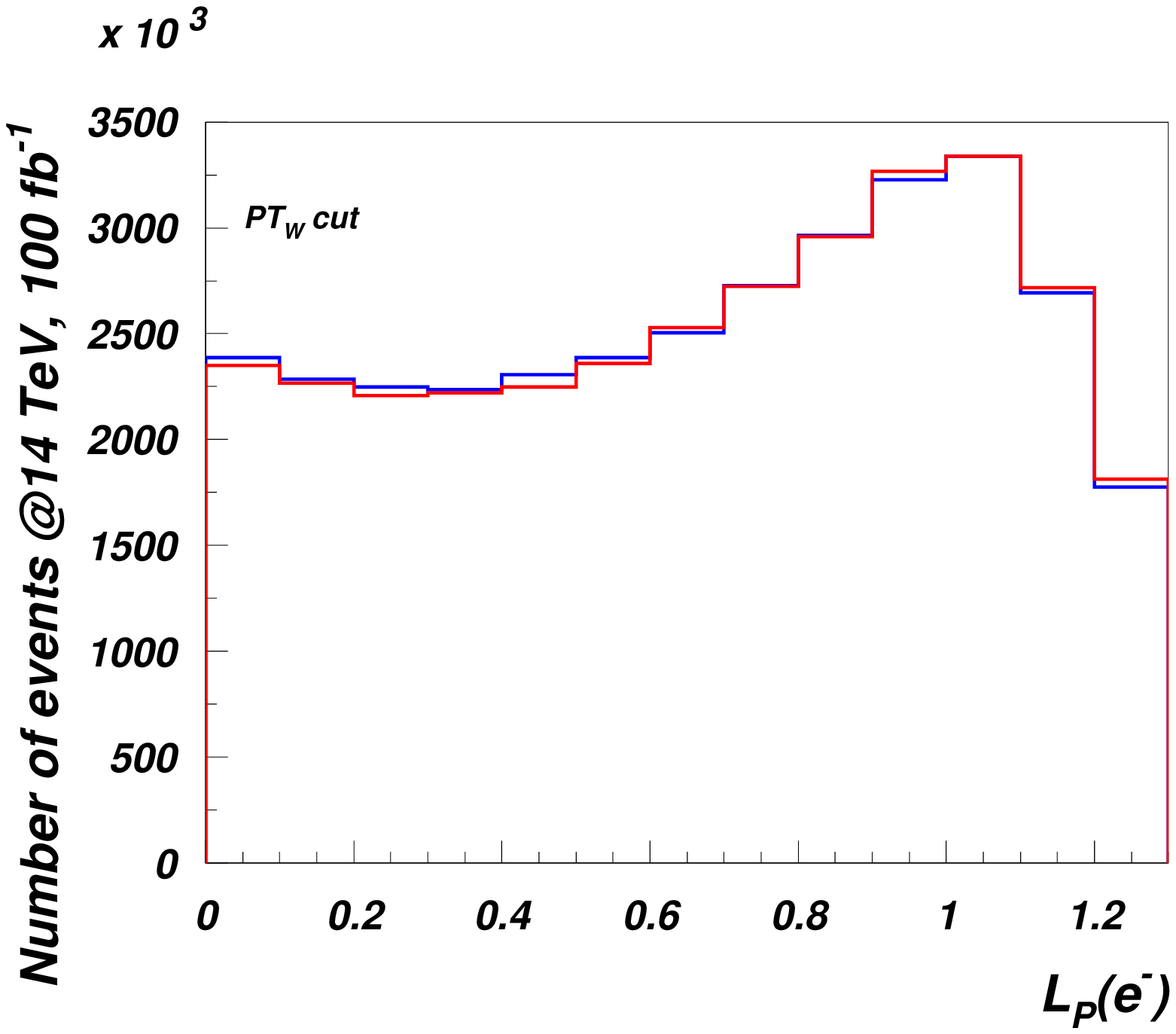}%
\includegraphics[width=0.5\textwidth]{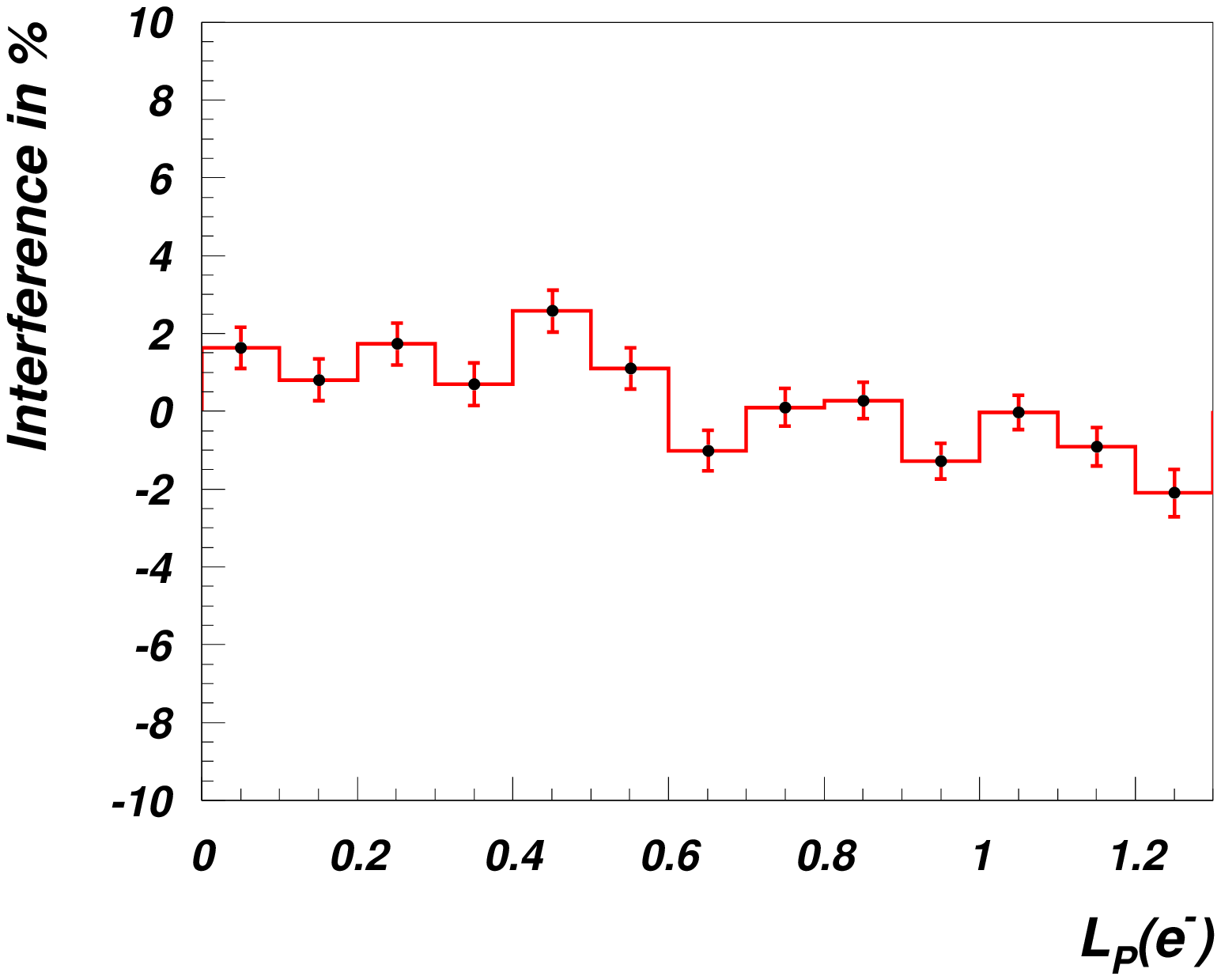}\\
\includegraphics[width=0.5\textwidth]{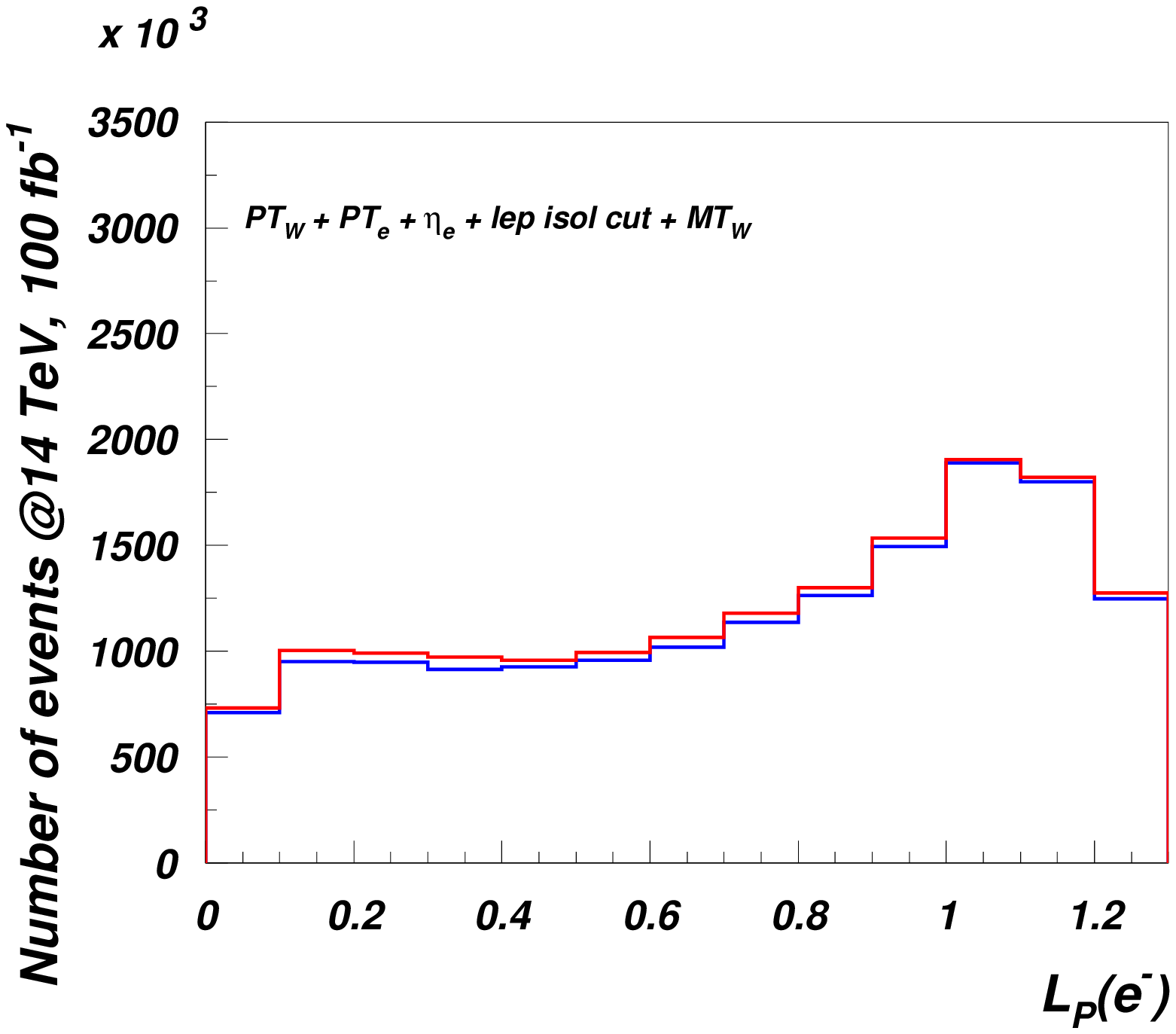}%
\includegraphics[width=0.5\textwidth]{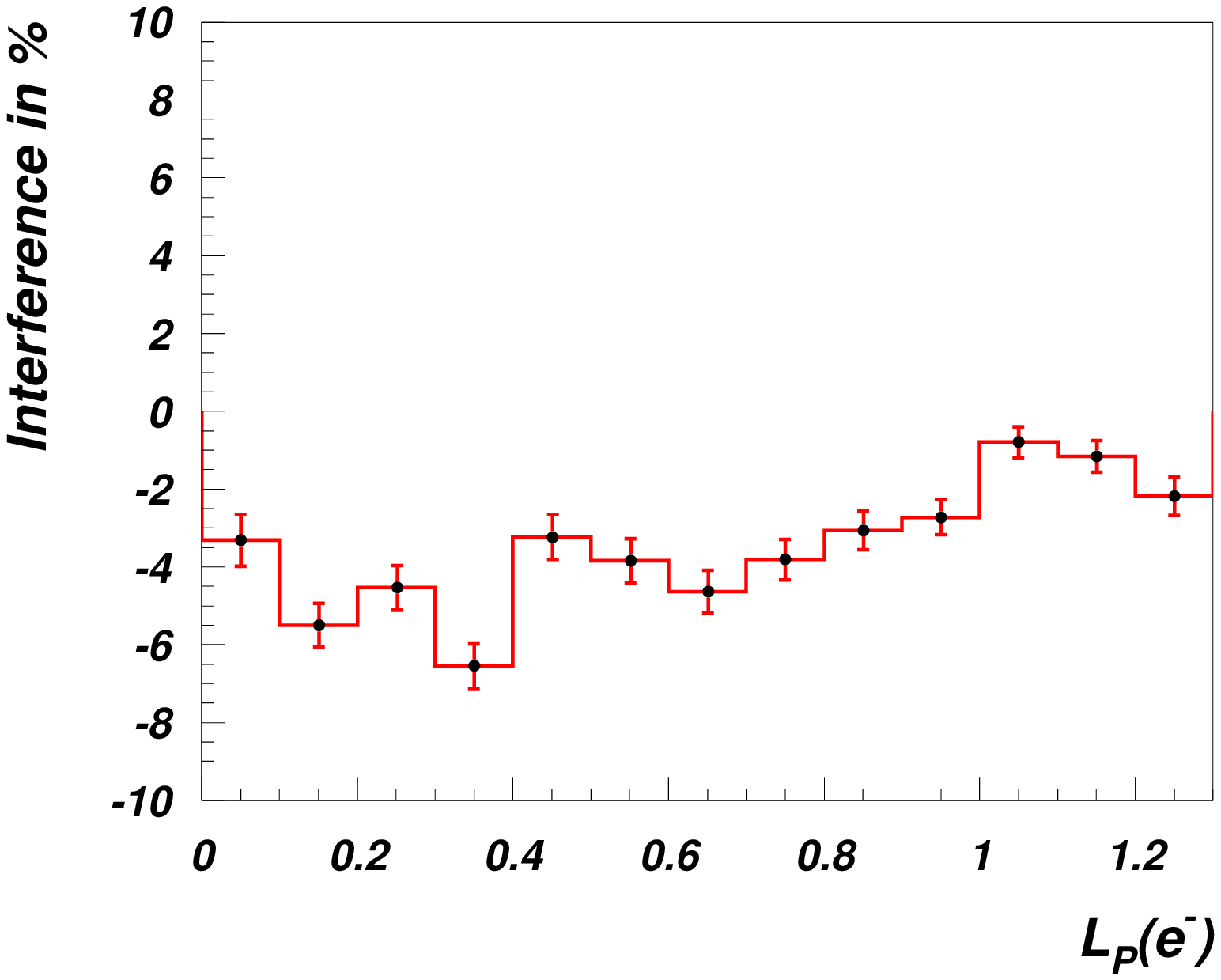}\\
\caption{\label{fig:interf-cuts_14}
Overall interference effect of $W$-boson polarisations
in  $d\sigma/dL_P$ \@14 TeV LHC with 100 pb$^{-1}$ after the consequent application of
the kinematic cuts from Table~\ref{tab:cuts}.
Left: distribution with(blue) and without(red) interference.
Right: the relative value of the interference versus $L_P$
including Monte-Carlo statistical error.
Upper: distributions after {\it Cut 1},
Lower: distributions after {\it Cuts 1-4}.}
\end{figure}

{
We have examined the sensitivity of the predicted differential cross-sections to different CTEQ PDF sets.
 The total cross-section differs by around 5\% between different sets, but the shape of the distribution in $L_P$ is fairly insensitive, i.e. if we renormalize so that the total cross-sections agree, the difference between the differential cross-sections in any
 $L_P$ bin is less than 2\% and therefore significantly lower than the effect of interference for most of the range of $L_P$ for the case of $W^-$ production.
In this paper we  used  the CTEQ6L set as this was the set used in ref.~\cite{Chatrchyan:2011ig}.
}

In Fig.~\ref{fig:interf-cuts_14} we present analogous results for  14 TeV LHC with 100 fb$^{-1}$ for 
the process
 $pp\to W^- + jet \to e^- jet+\ptmis$.
One can see that the shape of the interference for  the $L_P$ distribution does not change, however its relative size  decreases down to about 6\% for low $L_P$ bin. This is expected since at higher energies  the  $P_T(W)$
distribution is shifted to the higher end, which makes
 the approximation (\ref{approx}) more accurate - recalling
 that the interference effects cancel completely in the limit where
 eq.(\ref{approx}) becomes exact.

\subsection{LHC sensitivity to $g_A-g_V$ parameter space}

The CMS collaboration has studied sensitivity of the LHC to
measure the $W$-boson polarisation (\cite{Chatrchyan:2011ig}),
however this sensitivity does not provide a clear information about underlying theory.
 On the other hand, the fit of $L_P$ distribution as we show below
 could provide the  measurement of axial and vector couplings of  $W$-boson to fermions making  therefore a connection to the underlying Lagrangian.

\begin{figure}[htb]
\includegraphics[width=0.5\textwidth]{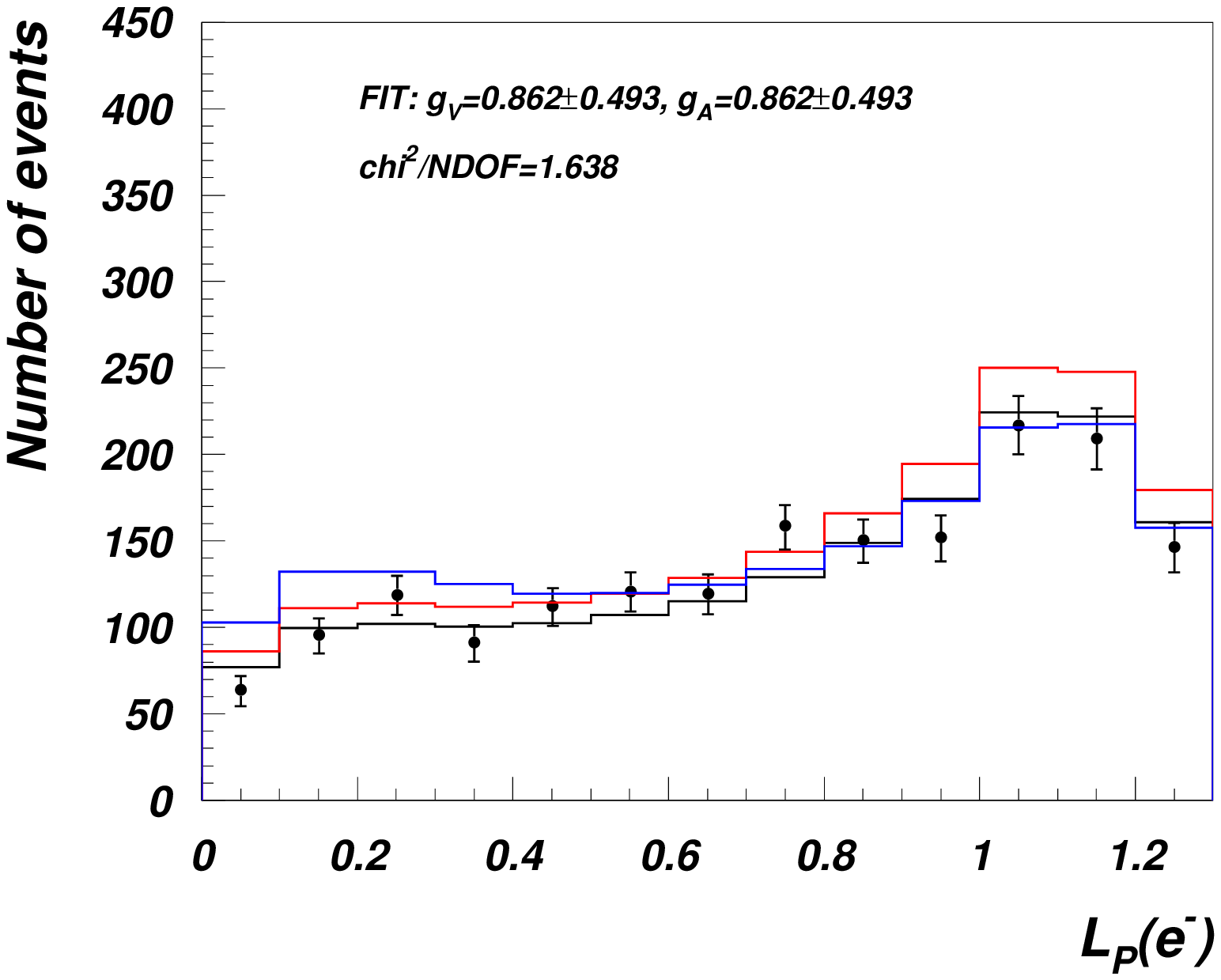}%
\includegraphics[width=0.5\textwidth]{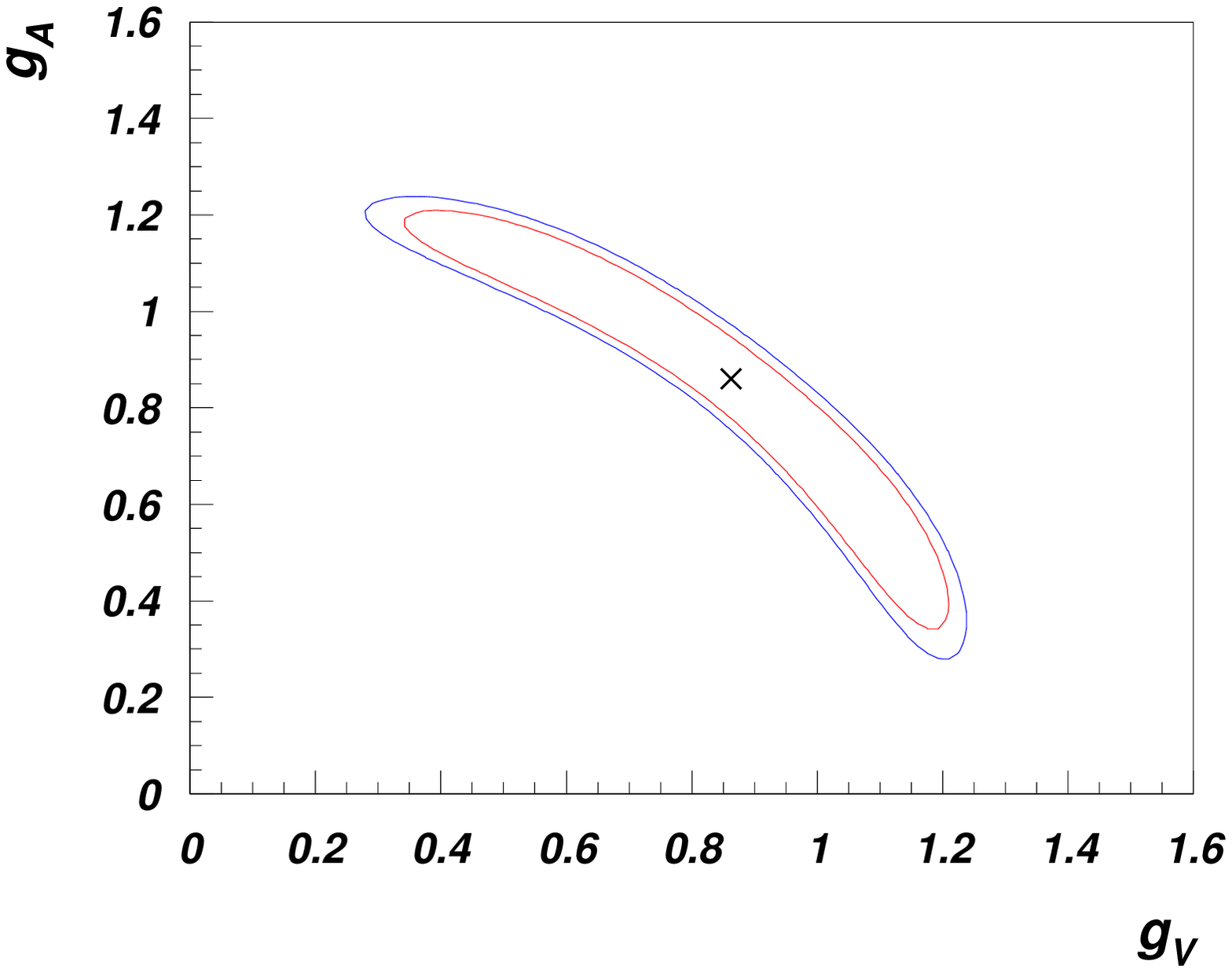}%
\caption{\label{fig:parton-level-fit} Left: Data (black dots)  fit (solid black)
 with   $L_P$ distributions (red is for $g_A=g_V=1$,
 blue is for  $g_A=0, g_V=\sqrt{2}$).
 Right: 67\% (red) and 95\% (blue) confidence level contours for
 the respective fit.
  Effects of calorimeter energy smearing, trigger efficiency as well as
effects are ignored.
}
\end{figure}

In this subsection we study the LHC sensitivity $g_A,g_V$ couplings of $W$-boson.
In Fig.~\ref{fig:parton-level-fit}(left) we present $L_p(e^-)$ distributions for
$g_A=g_V=1\mbox(SM)$ case (red line) as well as for $g_A=0, g_V=\sqrt{2}$
case represented by the blue line.
These distributions ignore the effects of calorimeter energy smearing, trigger efficiency as well as
effects of jet fragmentation but do include the application of Cuts 1-4.
Using these distributions we made an attempt to fit experimental data (including experimental errors) from~\cite{Chatrchyan:2011ig}  with $g_A, g_V$.
First of all, one can see that at such level of simulation  our $L_p(e^-)$ prediction describes data quite well -- it is only above data by about 10\%,
 {which is also indicated in the best values of $g_V$ and $g_A$:  
 $g_A^*=g_V^*=0.86\pm 0.5$.}
Also in the right panel of  Fig.~\ref{fig:parton-level-fit} we present 67\% and 95\%
confidence level contours for this fit. One can see that the sensitivity to the values of  $g_A,g_V$ is quite low.
One should note, that the contour plot in Fig.~\ref{fig:parton-level-fit}(right)
reflects a clear correlation between $g_A$ and $g_V$ parameters while the error
of the fit is reflected in the respective maximal variation of either parameter
$g_A$ or  $g_V$  at the 67\% confidence level.

In the next stage we have
performed fast detector simulation
using the PGS package~\cite{PGS}. At this level
we have taken into account effects of calorimeter energy smearing, trigger efficiency as well as
effects of jet fragmentation.

\begin{figure}[htb]
\includegraphics[width=0.5\textwidth]{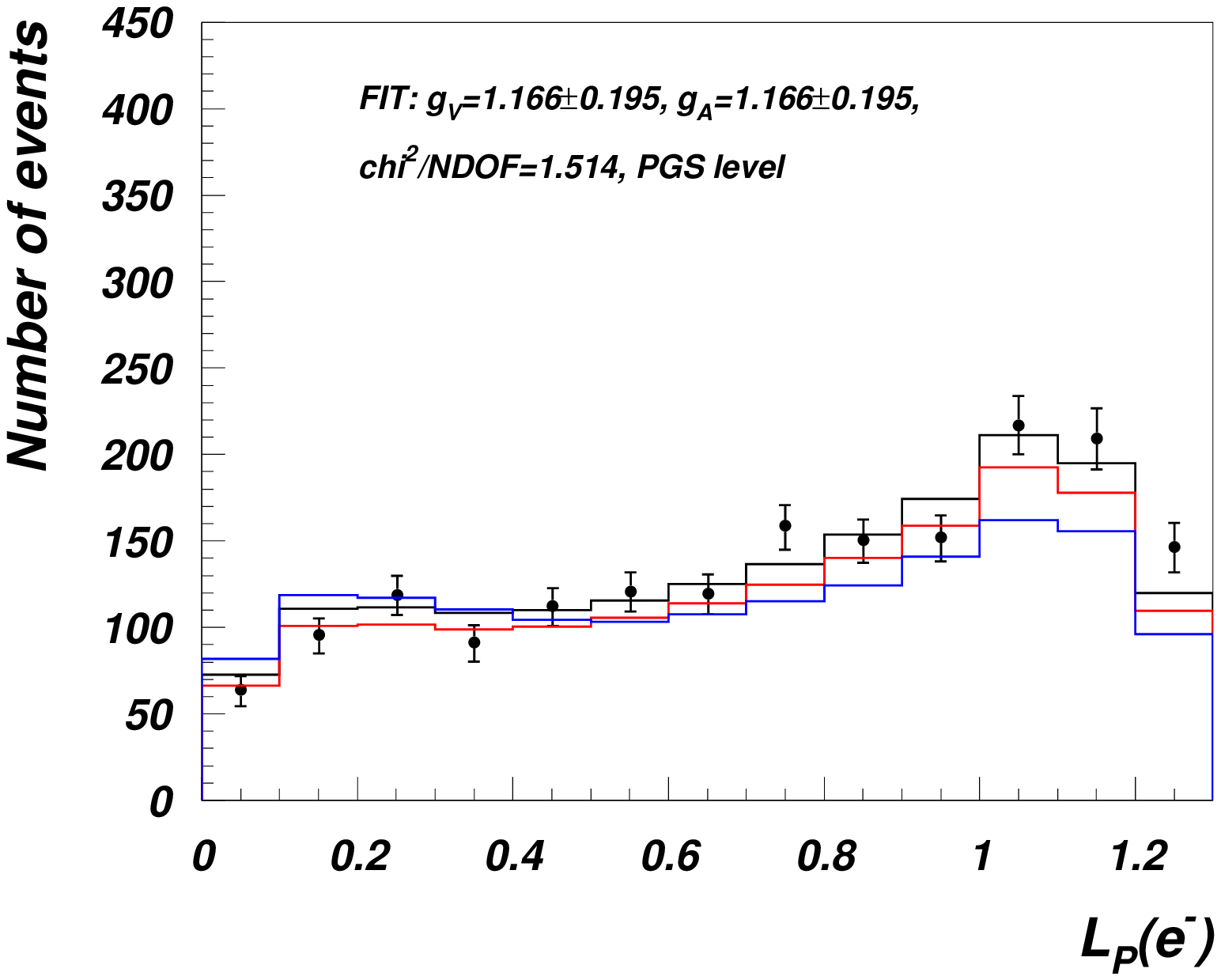}%
\includegraphics[width=0.5\textwidth]{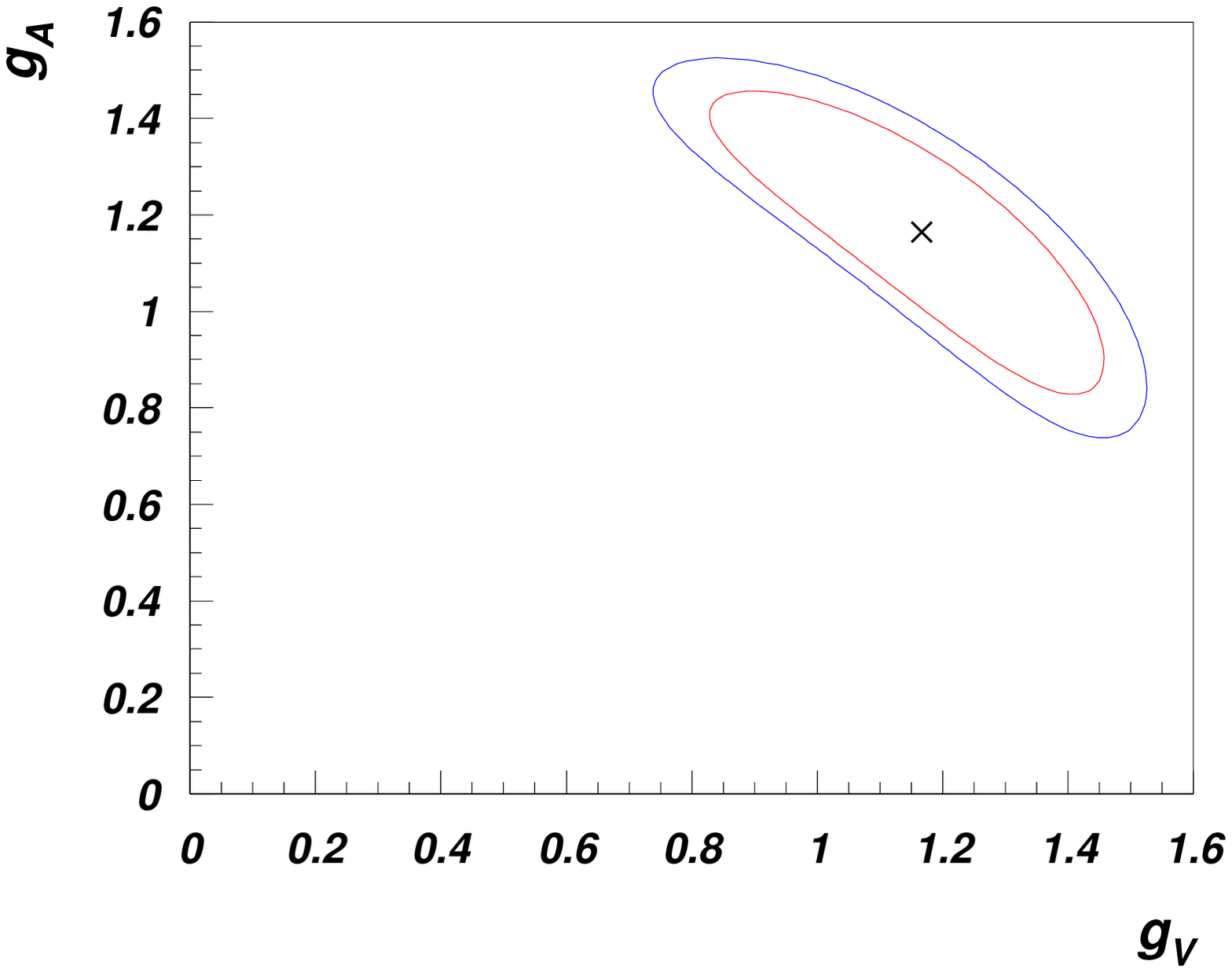}%
\caption{Left: Data (black dots)  fit (solid black)f
 with  PGS (fast detector level simulation) $L_P$ distributions (red is for $g_A=g_V=1$,
 blue is for  $g_A=0, g_V=\sqrt{2}$).
 Right: 67\% (red) and 95\% (blue) confidence level contours for
 the respective fit.\label{fig:pgs-fit}}
\end{figure}

In Fig.~\ref{fig:pgs-fit}(left) we present $L_p(e^-)$ distributions for
$g_A=g_V=1\mbox(SM)$ case (red line) as well as for $g_A=0, g_V=\sqrt{2}$
case represented by the blue line
analogous to the previous figure but after the use of the PGS package.

One can see that PGS slightly affect the shape of the $L_P$ distribution:
the distribution for $L_P\ge 1$ is somewhat more suppressed than the region 
of lower $L_P$. Noting that this region is dominated by low $p_T(W)$
and respectively low $p_T$ of its decay products and associated jets,
this indicates that under more realistic conditions the low $p_T$
events have slightly lower detector efficiency.\footnote{
Also, missing transverse momentum which is
determined from the vector sum of the transverse momentum of the lepton and the jets
(and used to construct $p_T(W)$ is more sensitive to PGS smearing effects
for low $p_T(W)$.}

One can see that now, the $g_A=g_V=1$  case
provides even better quantitative description of data,
which is indicated in the fitted values of  $g_A^*=g_V^*=1.17\pm 0.20$
as well as slightly better value of $\chi^2$.
The respective sensitivity in $g_A-g_V$ plane is
indicated in Fig.~\ref{fig:pgs-fit}(right).

\begin{figure}[htb]
\includegraphics[width=0.5\textwidth]{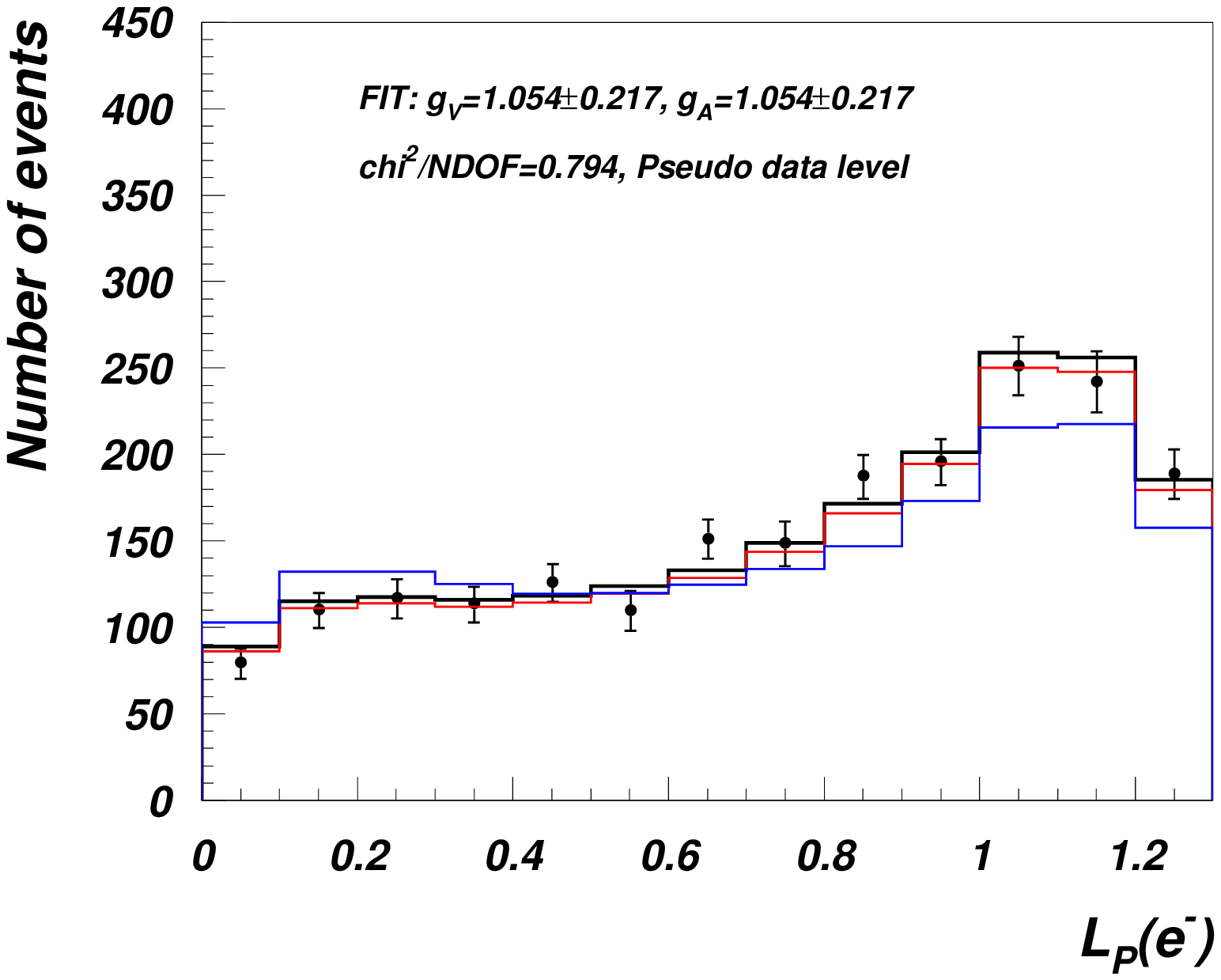}%
\includegraphics[width=0.5\textwidth]{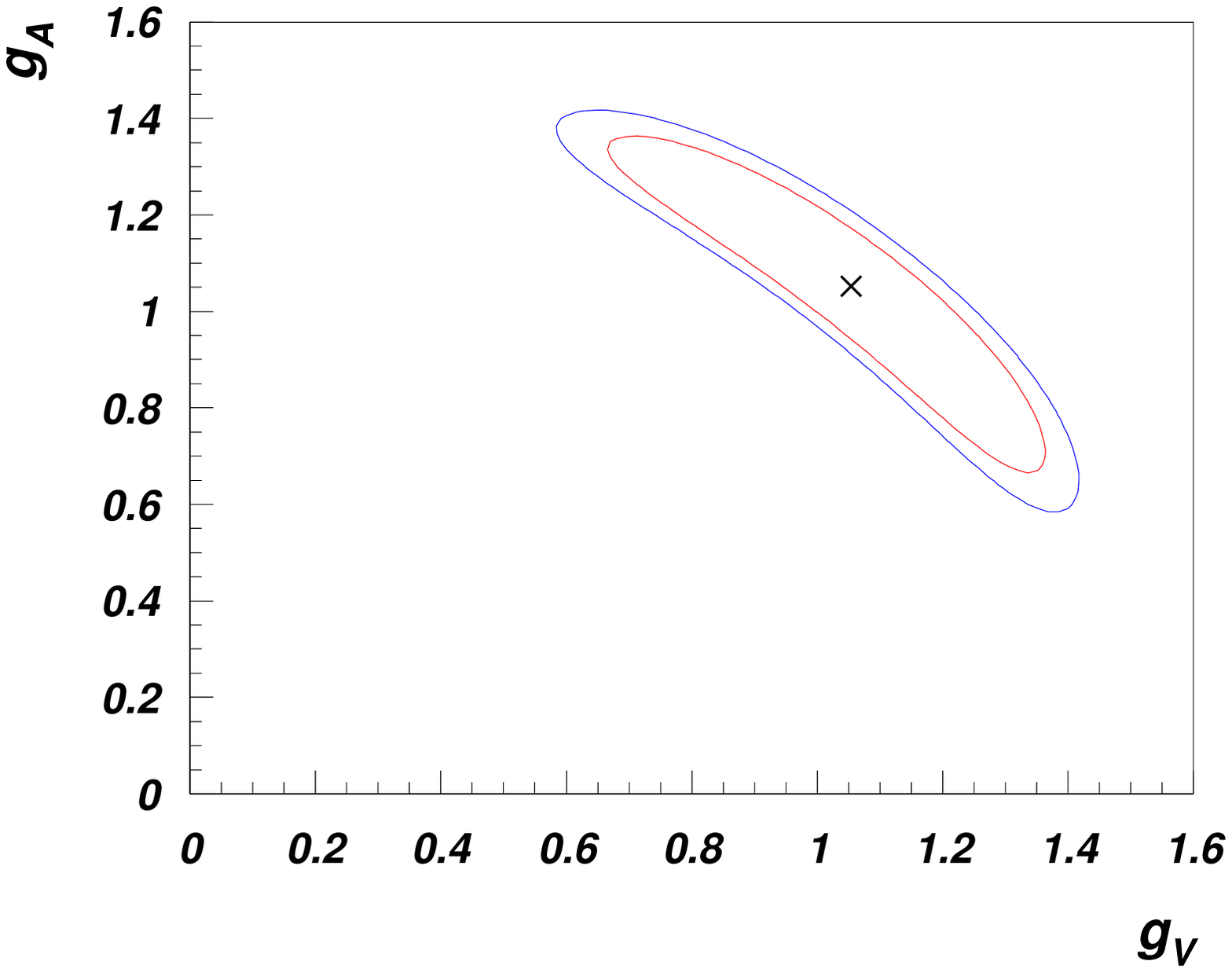}%
\caption{Left: ``Pseudo data" (randomised $g_A=g_V=1$ distribution -- black dots)  fit with ``parton level" $L_P$ distributions (red is for $g_A=g_V=1$,
 blue is for  $g_A=0, g_V=\sqrt{2}$).
 Right: 67\% (red) and 95\% (blue) confidence level contours for
 the respective fit.\label{fig:data-fit}}
\end{figure}

The full detector simulation would provide the ultimate level of data description,
therefore, assuming this,
we have randomised results for our $L_P$  distribution
around their mean values according to the Gaussian distribution
using available experimental errors. Using this ``pseudo data" level of analysis
we have estimated the LHC sensitivity to $g_V,g_A$ parameters
presented in Fig.\ref{fig:data-fit} in analogy to the previous ones.
One can see that in comparison with the approaches above,
we have achieved slighly better $\chi^2$ fit and the the sensitivity to
$g_V,g_A$ parameters  ($g_A^*=g_V^*=1.02\pm 0.16$).

For all three methods  of analysis above which are in a good agreement
between each other, one can see the LHC sensitivity to $g_V,g_A$ parameters
is still below the 30-40\% level. \footnote{ The panels on the RHS of figures 6 - 8  show that the errors on $g_V$ and $g_A$ are closely correlated, but the 
error on either one is as quoted above.}
This is related to the fact that even for the two extreme cases
of $g_A=g_V=1$,  and  $g_A=0, g_V=\sqrt{2}$
the respective shapes of $L_P$ distribution are not dramatically different.

With  the higher statistics which is already available for  7 and 8 TeV LHC and
much higher statistics which will be available at 14 TeV LHC, the main factor
which will  affect the  LHC sensitivity to the parameters $g_V, \, g_A$,
is clearly the systematic error which contributed about 50\%
(although this  might also be improved with higher statistics)
 in the CMS analysis at 7 TeV, 36 pb$^{-1}$.\footnote{One should also note that
at high luminosity pile-up effect will be  more dramatic
affecting the overall error.}
 
  So, one can expect the improvement of the sensitivity
to  $g_V,g_A$  by  not more than a factor of two given that the systematic error will stay the same.
This would mean that the accuracy to which LHC can measure  $g_V,g_A$ couplings 
is quite limited  and
would not probably better than  about 15\% level unless the systematic error is decreased.

\section{Conclusions}
In this paper we have discussed results of the CMS collaboration
on the sensitivity of the LHC to $W$ boson polarisation 
in the process  $pp\to W^\pm + jet \to e^\pm jet+\ptmis$
using the $L_P$ variable.

First of all we have shown that 
the differential cross section for
$pp\to W^\pm + jet \to e^\pm jet+\ptmis$  process
with respect to $L_P$ 
is sensitive to the interference between   different polarizations of $W$-boson
although it diminishes with the increase of  $W$-boson $p_T$.
The
interference effect is of the order of the  {statistical} experimental errors quoted in
ref.\cite{Chatrchyan:2011ig} and presumably significantly larger than the
current error bars that would be extracted from an analysis over the current (and future)
entire integrated luminosity.
{As explained above, the template method employed in 
ref.\cite{Chatrchyan:2011ig} does take account of the interference terms
in the form of a systematic 
uncertainty in the values of $f_+$, $f_-$ and $f_0$ quoted
with the interference terms taken to be within 10\% of
the Standard Model values.}
The results shown in Figs(2 - 4) show the entire interference
 effect, which is usually small but can be as large as 8\% in some bins.
Nevertheless, we suggest that if data with higher statistics were to be 
analyzed, it would be more informative simply to plot the measured differential cross-section with respect to $L_p$ and compare this with a MC simulation
that accounts directly for the interference, rather than attempt to extract values
for $f_+, \, f_-$ and $f_0$, or alternatively to include the interference coefficients, $g_{ij}$ into the fit procedure.

Although the sensitivity to polarisation of the  W-boson was shown to be
quite good -- of the order of 5\%
as we have estimated, the sensitivity to the underlying theory in terms of $g_V,g_A$ parameters 
is quite poor -- of the order of 30 - 40\%.
The  sensitivity of the LHC to   ($g_V,g_A$) plane via
the process  $pp\to W^\pm + jet \to e^\pm jet+\ptmis$
is limited by systematic error and is poor even in case of statistical error is very small.
The reason for this is the low sensitivity of the $L_P$ shape variation to  ($g_V,g_A$).
Even though, we should note that the variable
  $L_P$, suggested by CMS collaboration,
is a suitable  variable (and probably one of the best) 
to study LHC sensitivity to $g_V,g_A$ parameters since it is  directly connected to $\cos\theta^*$.

We have also found that only simulations at full detector  level would allow 
us to  fit data properly  and estimate
the LHC sensitivity   to  ($g_V,g_A$) parameter space.
As a result of this study we have created MC generator which linked to PYTHIA generator,
and  can be suitably used for study of the W-boson polarisation, effects of the interference and LHC sensitivity 
to  $g_V,g_A$ plane.
It is available upon request.
This generator can produce events in the generic LHE format which can be plugged into
the full detector simulation chain and used in the experimental analysis.

 Our observation of the fact that the  experimental  sensitivity
to W-boson polarization is very different and much higher than
the experimental sensitivity to  ($g_V,g_A$) parameter space can be turned around and used
to identify possible new physics at the LHC.
If  the $g_V-g_A$ parameter space is accurately measured from some different process
and agrees with the SM predictions whilst the polarisation of the $W$-boson
is found to be different from the SM expectations, this would be a  clear indication that
deviation of  $W$-boson polarisation comes {\it not} from  the  ($g_V, g_A$) couplings but from different sector, such as, for example new particles with different spin statistics
which would happen in case of supersymmetry.

Finally,  we would like to note, that, a full NLO corrections to this process
 have been have been calculated, the QCD corrections 
  in refs.\cite{Mirkes:1992hu,Bern:2011ie}
 and the  electroweak corrections in ref. \cite{Dittmaier:2003ej},
 but these results have not presented in terms of the variable $L_p$. The NLO
 corrections to the cross-section $d\sigma/dL_p$ would be of great interest
 and could well affect the results of this analysis.


\acknowledgments
We would like to acknowledge support from 
STFC Consolidated grant ST/J000396/1
as well as partial  support from
the NExT institute, a part of SEPnet.
We also thank Oliver Buchmueller, Jad Marrouche and Paraskevas 
Sphicas
for a very  useful discussions and clarifications.


\bibliographystyle{JHEP}
\bibliography{bib}
\end{document}